\documentclass[12pt,preprint]{aastex}

\shorttitle{Chemo-dynamical history of the CB17 core}
\shortauthors{Pavlyuchenkov et al.}

\begin{document}
\title{CB17: Inferring the dynamical history of a prestellar core with chemo-dynamical models}

\author{Ya. Pavlyuchenkov}
\affil{Max Planck Institute for Astronomy,
K\"onigstuhl 17, D-69117 Heidelberg, Germany}
\email{pavyar@mpia.de}

\author{D. Wiebe}
\affil{Institute of Astronomy of the Russian Academy of
Sciences,
48, Pyatnitskaya Str., Moscow, 119017, Russia}
\email{dwiebe@inasan.ru}

\author{R. Launhardt}
\affil{Max Planck Institute for Astronomy,
K\"onigstuhl 17, D-69117 Heidelberg, Germany}
\email{rlau@mpia.de}

\and

\author{Th. Henning}
\affil{Max Planck Institute for Astronomy,
K\"onigstuhl 17, D-69117 Heidelberg, Germany}
\email{henning@mpia.de}

\begin{abstract}
We present a detailed theoretical study of the isolated Bok globule CB17
(L1389) based on spectral maps of CS, HCO$^+$, C$^{18}$O, C$^{34}$S, and H$^{13}$CO$^+$
lines.  A phenomenological model of prestellar core evolution, a
time-dependent chemical model, and a radiative transfer simulation for
molecular lines are combined to reconstruct the chemical and
kinematical structure of this core. In addition we investigate the influence
of various physical factors on molecular line profiles. It is
shown that the intensity of the external UV field, the probability for
molecules to stick onto dust grains, the core age, and the rotation
velocity all significantly affect the molecular line spectra.
Due to this influence, the asymmetry of optically thick lines allows to remove
the ambiguity between the sticking probability and the core age.
We demonstrate that these parameters are well constrained, when results of
the modeling are compared to observations in multiple lines of sight through the core.
We developed a general criterion that allows to quantify the difference between
observed and simulated spectral maps. By minimizing this difference, we
find that very high and very low values of the effective sticking probability
$S$ are not appropriate for the studied prestellar core. The
most probable $S$ value for CB17 is 0.3--0.5. The spatial distribution of the
intensities and self-absorption features of optically thick lines is
indicative of UV irradiation of the core. By fitting simultaneously
optically thin transitions of C$^{18}$O, H$^{13}$CO$^+$, and C$^{34}$S as well as
optically thick transitions of HCO$^+$ and CS, we isolate the model that
reproduces all the available spectral maps to a reasonable accuracy
and, thus, represents a good approximation to the core  chemical and
kinematical structure. The line asymmetry pattern in CB17 is  reproduced
by a combination of infall, rotation, and turbulent motions  with
velocities $\sim0.05$ km s$^{-1}$, $\sim0.1$ km s$^{-1}$, and $\sim0.1$ km s$^{-1}$, respectively.
These parameters corresponds to energy ratios $E_{\rm rot}/E_{\rm grav}\approx0.03$,
$E_{\rm therm}/E_{\rm grav}\approx0.8$, and $E_{\rm turb}/E_{\rm grav}\approx0.05$
(the rotation parameters are determined for $i=90^\circ$).
The chemical age of the core is about 2 Myrs. In particular, this is indicated
by the central depletion of CO, CS, and HCO$^+$. On the other hand, the depletion
is not strong enough to show up in intensity maps as a ring-like pattern.
Based on the angular momentum value, we argue that the core is
going to fragment, i.e., to form a binary (multiple) star.
\end{abstract}


\section{Introduction}

Of all the diverse steps of the star formation process, the prestellar stage is
supposed to be the most quiescent phase.
Starless cores are nearly isothermal, have more or less regular shapes, and lack prominent kinematic features,
like disks and outflows. Among those cores, Bok globules are especially attractive for studies, as they
are relatively isolated from a confusing surrounding material and, thus, represent
`clean' examples of the prestellar chemical and dynamical evolution.
Numerous embedded and isolated starless cores have been subjects of chemical and kinematical
studies during recent years. As we generally want to know if these cores are going to become
stars eventually or, in other words, if they are
not only starless but really prestellar, these studies are mainly directed toward the search
for infall signatures. A commonly used collapse indicator is the characteristic blue-red
line asymmetry of optically thick lines \citep{evansaraa}. Usually, the central
spectrum of a core is utilized to demonstrate that this asymmetry is present.

However, if one wants to investigate the kinematics in more detail, it is necessary
to observe and to analyze spectra at locations offset from the center of a core.
1D spectral cuts or 2D spectral maps are employed in various ways.
Spectral maps for several molecules have allowed \cite{taf1998}
to detect an extended infall in the now famous L1544 prestellar core. This study has later been followed by
\cite{w99}, who have used N$_2$H$^+$ spectral maps to investigate the infall in the very center
of the core. Mapping surveys for infall motions in a number of starless cores have been carried out
by \cite{greg2000} in lines of HCO$^+$ and by \cite{lee2001} in lines of CS, CO, and N$_2$H$^+$.

Another goal of investigating spectral maps is the study of rotation of starless cores,
where both red and blue asymmetric line profiles are present.
E.g., \cite{kc1997} have studied the rotation in fifteen
starless Bok globules based on CO maps, while \cite{pavshust} have used HCO$^+$ spectral maps,
obtained by \cite{greg2000}, to determine the average rotation speed (as well as the shape
and the infall speed) for the L1544 core. \cite{pavshust} assumed that the core collapses
from the initial rigid rotation state so that its total angular momentum remains constant.
The spatial separation of red and blue asymmetric HCO$^+$
line profiles in the maps of the L1689B prestellar core has been reproduced with
a 3D molecular line radiation transfer model
by \cite{redman2004}. They showed that the asymmetry in this core can be explained
under the assumption of solid-body rotation in the core center. At a later evolutionary stage,
the rotation and infall profiles for the very young protostar IRAM~04191 have been determined
by \cite{belloche2002} from 	CS spectral maps.

A more comprehensive analysis of spectral maps is also possible. \cite{b68pulse} interpreted
the alternating asymmetry pattern in CS spectral lines observed in B68 as a signature of
simultaneous inward and outward motions having both radial and non-radial modes.
\cite{taf2004} performed a similar study of the L1498 and 1517B prestellar cores. Using
N$_2$H$^+$ and NH$_3$ spectral maps they investigated velocity gradients across the core faces
and found them indicative of a velocity pattern more complicated than just infall
or rotation. They suggested that this pattern is associated with asymmetric gas motions
resulting from residual core collapse.
These results indicate clearly that, given the overall quest for infall in starless
cores, a great care must be taken to avoid confusion between infall and other bulk motions.
They also stress that detailed kinematic information on starless cores is necessary
to draw any conclusion about their dynamical and evolutionary state.

The spectral maps are not the only ingredient needed for an analysis of the starless
core kinematics. To serve this purpose they have to be complemented by some knowledge on the
core chemical composition. The uniform abundances of tracer molecules over the core
are now confidently rejected both on observational \citep[e.g.,][]{sysdiff} and theoretical
\citep{rawyates} grounds. It is now customary to reproduce observations with some simplified
representation of a radial abundance profile (like a step function, for
example). However, \cite{lee2005} showed that simple empirical distributions may not be good indicators of the
real chemical structure in collapsing cores. But if they would be, these simple distributions
do not contain information about the dynamical and chemical history of the core. \cite{rawyates} argued
that only multiple line-of-sight observations coupled to detailed chemical
models are able to shed light on prestellar evolution.

This is the approach we adopt here, utilizing detailed spectral maps in a number of transitions
for the CB17 core. This core is located at the south-east
edge of a small isolated, and slightly elongated globule at a distance of about 300~pc.
Several authors have studied its chemical composition, but the most extensive investigation
has been performed by Turner and co-authors \citep{TurnerIV,TurnerVII,TurnerVIII,TurnerIX}. They argued
that the chemical composition of this core is in many respects unusual, with overabundant
HCO$^+$, N$_2$H$^+$, HC$_3$N, and some other species. However, it must be noted that their
study has been focused on small translucent clouds. When compared to other dense cores, the
CB17 core shows typical column densities of N$_2$H$^+$ \citep{benson1998,cas2002} and ammonia
\citep{lemme1996}.

Using molecular line maps, we study in this paper abundances of
CO, HCO$^+$, and CS as a function of distance from the core center.
We choose a detailed chemical model to compute the time-dependent abundances
at a number of locations within the core, assuming its spherical symmetry. Both static
and dynamically evolving configurations are considered. These abundances are then used to simulate
the spectral maps that are directly compared to the observed line profiles.

The attempt to reproduce the observed state of the core as an outcome of the time-
and depth-dependent chemical model is hampered by the presence of many unknown
or poorly known parameters. At first sight, it may seem they can be adjusted almost
arbitrarily to get the best agreement with observations. However, this is not exactly the case
when muliple lines of sight are analyzed. The most important parameters for the chemical model
are the probability $S$ for species to stick to dust grains and the intensity $G$ of the UV~field
that illuminates the core, as they control abundances in the inner and outer regions of a core,
respectively. Two remarks should be made about these parameters. First, the sticking
probability $S$ represents an effective value which is valid as long as there are no
desorption mechanisms other than those considered in the paper (thermal desorption,
cosmic ray desorption, photodesorption). In the case that some powerful evaporation
mechanism is missing in the model, the best-fit $S$ value would be higher than that
quoted in the Abstract. Second, rates of photoreactions can be different from the rates used in the model
if the spectral properties of the UV field in the core vicinity differ significantly from the
average interstellar field.

We consider a range of these parameters and demostrate that their values can be significantly
constrained if a comparison with observations is not limited to just a central spectrum.
Using a static model, we find the value of $G$ that allows to reproduce
the observations most reasonably. This value is then used as an input for the dynamical model.
Instead of fixing a certain contraction law, we consider a range of {\em ad hoc\/}
solutions to check if the dynamical history of a core leaves some observable imprint
on its chemical structure and emergent line profiles. With the dynamical model we estimate
the sticking probability, the age of the core, the infall velocity, and the rotation velocity.
In particular, we show that the asymmetry of the line profiles can be used to relax the
ambiguity between the sticking probability and the core age.

The structure of the paper is as follows. In \S~2 we describe the chemical model and the formalism
that is used to simulate the core dynamical evolution. The radiation transfer model, the criterion
for the comparison with observations, and the used observational data are also presented in \S~2.
Results of the chemical modeling
for the static core model are given in \S~3. The dynamical models with and without rotation are presented in
\S~4. Our results are discussed in \S~5 and summarized in \S~6.

\section{Model}

In general, we expect that the dynamical evolution of a prestellar core is related
to its chemical evolution via the time-dependent density, extinction etc.
Thus, we want to address the question if simulations of the chemical
evolution are a good tool to get insights into the dynamical history of the core.
For that, we consider a model of a collapsing and rotating
cloud, which provides us with the time-dependent density and velocity fields.
This model is combined with an appropriate chemical model in order to find 
the representaion of the core that provides the best agreement with the observed spectral maps.
A straightforward way is
to parameterize the problem and then to use some criterion that quantifies the difference
between the observed and theoretical spectra. Given the large number of parameters,
we split the problem in three parts.

In a first step, we investigate the `chemical' parameters with a static configuration in
which the core has zero regular velocity, and the
density distribution stays constant with time and is equal to the observed current state
(see \S~\ref{statmod}). Ignoring the dynamical evolution, this model represents a
convenient way to study purely `chemical' factors which must be taken into account, when
analyzing molecular line spectra. Specifically, the static model is used to constrain
the intensity $G$ of the UV field in the vicinity of the core.

In a second step, the best-fit $G$ value is used in a 1D collapse model, in which
a phenomenological way is adopted to describe a time-dependent density structure.
This technique is by no means hydrodynamical in nature. However, it allows to
probe a wider range of collapse regimes than more physical approaches taken by, e.g.,
\cite{shem} or \cite{aikawa2005}. At this step we consider only the
central spectrum as it is not affected by rotation. This analysis permits the
selection of a collapse mode that is the best in reproducing the central line
asymmetry.

\begin{figure}
\includegraphics[width=0.7\textwidth]{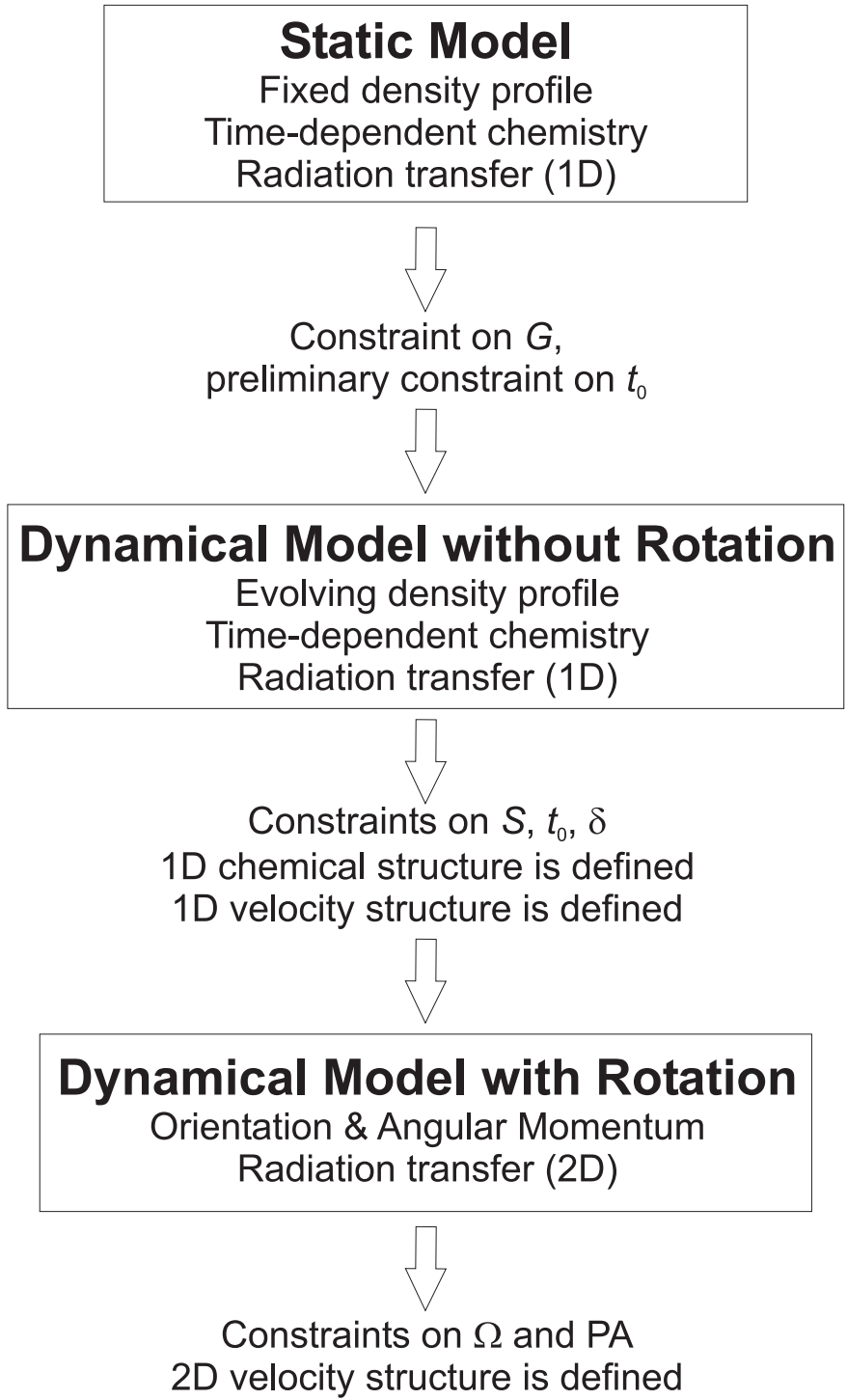}
\caption{The flowchart of the technique used in the paper.}
\label{flchart}
\end{figure}

In a third step, the best-fit dynamical model is used as an input for the model with both
collapse and rotation. With this model we determine the rotation velocity
and the position angle. A flow chart of these steps is presented in Figure~\ref{flchart}.
\clearpage
\clearpage
In the following, we describe both the static and non-static models in detail, together
with the adopted models of the chemical evolution and the line radiation transfer.
A criterion is presented that is used to compare modeled and observed spectral line maps.
For this comparison, we consider the optically thin transitions C$^{34}$S(2-1), H$^{13}$CO$^+$(1-0), and
C$^{18}$O(2-1) as well as the optically thick transitions CS(2-1), HCO$^+$(1-0), for which
most complete spectral maps are available. As we have one
transition per isotopomer, hereafter we omit the transition designation.

Observations of various molecular emission lines towards CB\,17 for isotopes and transitions of CO, CS, and HCO$^+$ were performed in June 1993 and October 1996 using the IRAM 30-m telescope. Final calibration and data reduction were done with the CLASS software, which is part of the GILDAS\footnote{The GILDAS working group is a collaborative project of Observatoire de Grenoble and IRAM, and comprises: G. Buisson, L. Desbats, G. Duvert, T. Forveille, R. Gras, S. Guilloteau, R. Lucas, and P. Valiron.} package. Since different nominal center positions as well as $v_{\rm LSR}$\ were used in different years, and some of the maps were done with a smaller spatial sampling for the central part of the dense core than for the outer regions, all spectra of a given line were combined and resampled on a regular spatial and frequency grid using a gaussian beam of HPBW and channel width listed in Table~\ref{tbl-2}. The reference position (0,0) for all data and maps in this paper is RA\,=\,04:00:32.90, DEC\,=\,56:47:52.0 (B1950). These observations represent part of an extensive study of the CB17 core in many molecules and transitions which will be described in full detail in another paper (Launhardt et al., in preparation).
\clearpage
\begin{table} 
\caption{Spectral line parameters at peak position (-1\arcsec,+8\arcsec) 
\label{tbl-2}} 
\begin{tabular}{llll} 
\hline 
Line & HPBW, arcsec & $\Delta v_{\rm chann}$, m\,s$^{-1}$ & $v_{\rm LSR}$, km\,s$^{-1}$ \\ 
\hline 
C$^{18}$O\,(2--1)          & 11 & 100 & -4.73 \\ 
CS\,(2--1)                 & 25 &  70 & -4.68 \\ 
C$^{34}$S\,(2--1)          & 25 &  65 & -4.59 \\ 
HCO$^{+}$\,(1--0)          & 27 &  65 & -4.67 \\ 
H$^{13}$CO$^{+}$\,(1--0)   & 29 &  65 & -4.64 \\ 
\hline 
\end{tabular} 
\end{table} 
\clearpage
To compare models and observations we need to define
velocity shift for modeled spectra, i.e. $V_{\rm LSR}$, which
represents velocity of a model core with respect to an observer.
Obviously, this value should be the same for all the studied lines.
However, gaussian fit to the observed profiles toward the center of the
core produces different $V_{\rm LSR}$ values for various lines
(between $-4.59$ and $-4.73$~km s$^{-1}$, Table~\ref{tbl-2}).
Within the framework of our approach
we cannot explain this scatter of $V_{\rm LSR}$ for central profiles. On
the other hand, differences in observed $V_{\rm LSR}$ values may be caused not by
a presumably complex structure of the core but rather by somewhat incorrect
values of rest frequencies for the studied transitions.

For simplicity, in our calculations we adopt $V_{\rm LSR}=-4.7$~km s$^{-1}$ for HCO$^+$(1-0),
CS(2-1), and C$^{18}$O, and $V_{LSR}=-4.6$~km s$^{-1}$ for
H$^{13}$CO$^+$(1-0), C$^{34}$S(2-1).

\subsection{Morphology and Dynamics}

\subsubsection{Static Core}
\label{statmod}
The density structure of a prestellar core can generally be approximated by
\citep[e.g.,][]{sysdiff}
\begin{equation}
  n({\rm H_2})=\frac{n_0}{1+(r/r_0)^{\beta}},
  \label{dens}
\end{equation}
where $n_0$ is the central density, $r_0$ is the radius, where $n$ has dropped
to half the central density (i.e., the size of an inner density plateau), and $\beta$ is the power
law index which describes the density fall-off in the outer parts of the
core. We use equation (\ref{dens}) to describe the observed density
profile of the CB17 core with parameters $n_0=5.4\cdot10^5$~cm$^{-3}$, 
$r_0=3000$~AU, and $\beta=2.2$ as derived in Launhardt et al. (in preparation). In addition, we use
an outer cut-off radius which is taken to be
$2.5\cdot10^4$~AU, the estimated radius of the CB17 core. The gas
temperature in the CB17 core is assumed to be uniform and equal to
$T=10$~K.

In the static model the core evolves only chemically, with an evolutionary
time $t_0$. In physical terms,
this model corresponds to the situation where the core forms almost instantly
(with the formation timescale significantly shorter than the chemical
timescale) and then stays static for its entire lifetime.
In the absence of a systematic velocity, the width of line
profiles is determined by microturbulent  velocity $V_{\rm turb}$, which is taken
to be 0.15~km s$^{-1}$ over the entire core in order to reproduce the observed
line width.

\subsubsection{Collapsing Core}
\label{dynmodsect}

The static model does not provide a complete description
of a typical starless core, as many such cores (including CB17) show
clear signs of internal dynamics. Molecular abundances certainly do not
only depend on the core density and `chemical' age, but rather represent the
result of its entire previous chemo-dynamical evolution.
This is why almost all recent chemical studies of prestellar cores are
coupled in some way or another to dynamical models \citep[e.g.,][]{shem,aikawa2005,lee2005,flower}.
In this paper we do not stick to any particular dynamical solution. Instead we choose
a phenomenological approach that allows to describe
the core evolution with only a few parameters. 

\clearpage
\begin{figure}[tbp]
\centering
\includegraphics[width=0.48\textwidth]{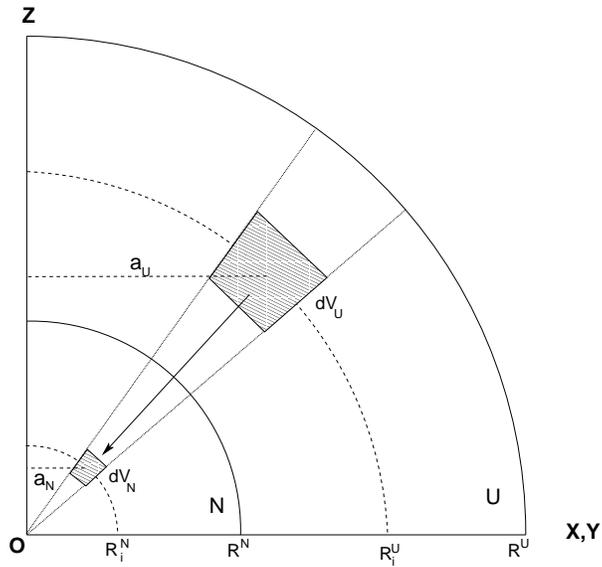}
\caption{Phenomenological model of the starless core evolution. $R^{\rm U}$
is the radius of the initially uniform core, $R^{\rm N}$ is the radius of 
the non-uniform (observed) core. The volume element $dV_{\rm U}$ during
collapse evolves to the volume element $dV_{\rm N}$.}
\label{scheme}
\end{figure}
\clearpage

As in the static model, we assume that the non-uniform (`N') density distribution 
(\ref{dens}) represents the current, observed state of the core. In this configuration,
the model core is divided into 48 concentric shells of equal width, having
radii $R_i^{\rm N}$.
In the initial uniform (`U') state the model core is spherically
symmetric and has the constant initial density
$n_{\rm U}=5\cdot10^3$~cm$^{-3}$ (Figure~\ref{scheme}), which is close to the typical
gas density in the vicinity of the core. Our results are not sensitive to small variations
of this parameter. Locations of shell
boundaries in the `U'-configuration are determined by the requirement that
each shell has the same mass as in the final state, i.e.
the initial radius of each shell $R_i^{\rm U}$ is uniquely defined by the
assumed $n_{\rm U}$. The time scale for
the evolution of the system, needed for configuration `U' to evolve to
configuration `N', is $t_0$. At any moment $t$ the location of the $i$th shell
is defined by
\begin{equation}
R_i(t)= R_i^{\rm U}- W_i\left(\frac{t}{t_0}\right)^{\delta}
\label{law}
\end{equation}
where $W_i$ is defined by the condition $R_i(t_0)= R_i^{\rm N}$, so that
$W_i=R_i^{\rm U}-R_i^{\rm N}$. Using equation (\ref{law}), we compute the density of each
shell as a function of time and use this in the chemical model. The radial velocity
of shell $i$ at time $t_0$ is
\begin{equation}
V_i(t_0)=-\delta\,\frac{W_i}{t_0},
\label{vel}
\end{equation}
Thus, the velocity only depends on the $\delta/t_0$ ratio, given the initial density is fixed.
The power law index $\delta$ in equation (\ref{law}) allows to
describe different regimes of the core collapse without going into
details of an underlying physical model.

\begin{figure}[tbp]
\centering
\includegraphics[width=0.9\textwidth]{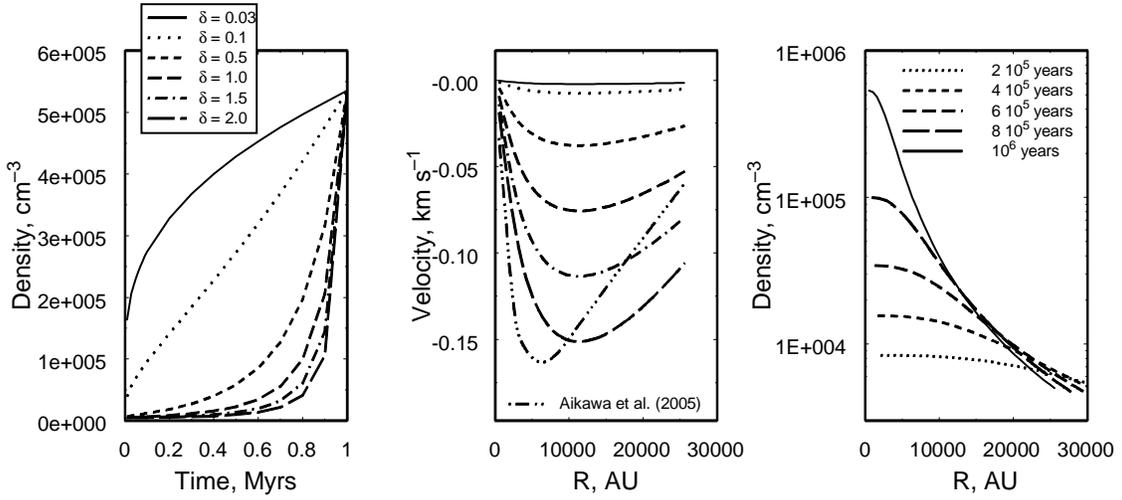}
\caption{Dynamical structure of the model core. (left) Time evolution of the central number density
for $t_0=10^6$~years and different values of $\delta$.
(middle) Radial velocity profiles for different values of $\delta$
at the end of evolution ($t_0=10^6$~years) compared to the velocity
profile from \cite{aikawa2005} for their $\alpha=1.1$ and $t=1.15\times10^6$~years.
Same legend applies as in left panel. (right) The density profile evolution for 
$t_0=10^6$~years and $\delta=1$. The density distribution at the last time moment
corresponds to that inferred for the CB17 core.}
\label{dynmod}
\end{figure}

As an example, in Figure~\ref{dynmod} we show plots for the central density evolution and
the radial velocity profiles in models with $t_0=10^6$~years and various values
of $\delta$. A nearly linear growth of the central density corresponds to
$\delta=0.1$. At smaller $\delta$, density accumulation decelerates with time;
at $\delta>0.1$ the density first stays almost constant, but then grows faster and
faster. The static model corresponds to $\delta=0$.

This model is similar in concept to other
generalized prescriptions given, e.g., by \cite{wwt} or \cite{myersgen}.
It is also close (for $\delta\approx2$) to the destabilized
Bonnor-Ebert sphere collapse model used by \cite{aikawa2005}, which is also shown
in Figure~\ref{dynmod}. The velocity
profile in our model is shallower and shows less tendency to peak at the
near-core region. This is caused by a different choice of initial conditions,
specifically, by the uniform initial density distribution.

This prescription for the core contraction may not
 satisfy the momentum equation strictly and is therefore
potentially inconsistent. However, our intention here is more to demonstrate, that it is possible in principle to
distinguish different regimes of the core evolution using observations 
of molecular lines, rather than to give support to some self-consistent solution. Also, the dynamics of prestellar cores can be
influenced by magnetic fields and/or turbulence, so that the simplest form of momentum 
equation for spherical isothermal collapse may not be 
satisfied anyway. Using
this approach, it may be possible to find the extent to which the momentum
conservation in collapsing cores is affected by non-thermal supporting factors.
\subsubsection{Collapsing Core with Rotation}
\label{rotomega}

Even though the rotation of starless cores does not influence their
(early) dynamical evolution it affects molecular line profiles.
The CB17 core is likely to have a significant angular momentum, which shows up
as an alternating asymmetry pattern across the core face. 

We include rotation into the phenomenological model, assuming that a
toroidal element $dV_{\rm U}$ of a shell (Figure~\ref{scheme}),
initially located at a distance $a_{\rm U}$ from the rotation axis,
moves during collapse so that its angular momentum is conserved, i.e. that
the momentum is not redistributed over the core. This is a good approximation
as long as there is no magnetic braking or turbulent momentum transport. We assume that
the core initially rotates as a solid body with the angular velocity~$\Omega$,
so the azimuthal velocity at $t_0$ is
\begin{equation}
   V_{\phi}=\frac{\Omega a^{2}_{\rm U}}{a_{\rm N}},
\label{Vphi}
\end{equation}
where $a_{\rm U}$ and $a_{\rm N}$ are radii of a toroidal element in
configurations `U' and `N'. Each of the 48 shells is subdivided into 32 angular cells.
This model is used in the
following to estimate the angular momentum of the CB17 core and to
derive its spatial orientation.

\subsection{Chemical Model}
\label{chem-mod-sect}

The model for the chemical evolution of the core is
described by \cite{Wiebe2003} and \cite{Semenov2004}. We refer the
reader to these papers for more details. Here only the main features of the
model are summarized. The model is a time-dependent chemical model which includes
gas-phase reactions as well as the freezing-out of molecules onto dust 
grains and their desorption back to the gas-phase. For simplicity, all
grains are assumed to have the same radius of $10^{-5}$~cm. Surface
reactions are not taken  into account in the current study.

Gas-phase reactions are taken from the UMIST\,95 ratefile \citep{umist95}.
We consider the evolution of species containing H, He, C, N, O,
Mg, Na,  Fe, S, and Si atoms. For the cosmic rate ionization rate the standard
value of $1.3\times10^{-17}$~s$^{-1}$ is assumed. The UV-flux
for photo-reaction rates is expressed by the $G$ factor measured in
units of the average interstellar flux \citep{Draine1978}.
We are aware that the spectrum of the radiation field in star-forming regions
may differ quite significantly from the average interstellar spectrum.
The relative `hardness' of a spectrum affects the photoreaction rates. However,
the study of this effect is beyond the scope of the present paper. The CB17
core is quite isolated, so that there are no young massive stars nearby and no obscuring
molecular cloud is present. Thus, the spectrum of the radiation field should be similar to that
of the interstellar field.

As in our dynamical 
model the density profile evolves, the extinction is evaluated as a function
of time at each point of the flow as
\[
A_{\rm V}(r,t)=N_{\rm H}(r,t)/1.59\times10^{21}\,{\rm cm}^{-2},
\]
where
\[
N_{\rm H}(r,t)=\int\limits_r^Rn(H)\,{\rm d}r
\]
is the column density of hydrogen nuclei measured from the core
boundary $R$ to a point at the radius $r$. H$_2$ self- and
mutual shielding of CO and H$_2$ are taken into account using the results
obtained by \cite{Lee1996}.
No attempt is made to account for the thermal balance in the medium. The
model core is assumed to be isothermal at 10\,K. This assumption breaks
down at the core edge where the gas is heated by UV radiation. However
this moderate heating would not affect the chemical reaction rates
significantly. The same is true for the desorption from grains, as this region is
dominated by photo-desorption.

Neutral species other than H$_2$ and He are assumed to stick to dust grains
with the same probability $S$ which is one of the parameters of our study.
In addition to photo-desorption, thermal desorption and cosmic ray
induced desorption are taken into account. Desorption energies are taken
from \cite{hh93}.

At $t=0$ all elements are present in atomic form with the only exception of
hydrogen which is entirely bound in H$_2$ molecules initially. The `low metal'
initial abundance set from \cite{Wiebe2003} is used.
The entire core is chemically uniform initially. After the onset of
collapse, in the adopted Lagrangian description each gas parcel moves
and evolves independently with the density varying according to the
adopted dynamical prescription.

In our study we vary two parameters for the chemical 
model, namely, the strength of the external UV field $G$ and
the sticking probability $S$, which regulate the
abundances of species in the outer envelope and in the core center,
respectively.

As a test of the model, we compare it to results obtained by \cite{aikawa2005}.
The closest match to their model with $\alpha=1.1$ \citep[see][for details]{aikawa2005}
and central density of $3\times10^6$~cm$^{-3}$
is given by our dynamical model with $\delta\approx2$ (Figure~\ref{dynmod}). For this comparison, we
set $G=1$, $S=1$, $A_{\rm V}=3$ mag at the core edge and included surface reactions.
Also, the branching ratio for the N$_2$H$^+$ dissociative recombination
from \cite{geppert} is taken into account.
In Table~\ref{comparaikawa} we compare column densities from this calculations
to those given in Table~2 by \cite{aikawa2005}. We present both straight
column densities toward the core center and the column densities convolved
with a $40^{\prime\prime}$ Gaussian beam (HPBW; the distance of 300~pc is assumed). Given the variety of assumptions
and the different underlying reaction sets (UMIST\,95 and the NSM), the agreement
seems to be reasonable.
\clearpage
\begin{table}
\caption{Molecular column densities for a representative dynamical model}
\label{comparaikawa}
\begin{tabular}{llll}
\hline
Species & \cite{aikawa2005} & This work & This work, with convolution \\
\hline
CO         & 1.9(17) & 6.3(16) & 7.1(16)\\
HCO$^+$    & 2.9(13) & 1.2(13) & 1.3(13)\\
HCN        & 8.0(14) & 3.6(14) & 2.3(14)\\
HC$_3$N    & 5.8(12) & 4.7(13) & 3.9(13)\\
NH$_3$     & 7.8(15) & 4.1(15) & 1.5(15)\\
N$_2$H$^+$ & 1.8(13) & 8.4(13) & 2.8(13)\\
CS         & 1.0(13) & 4.5(13) & 4.5(13)\\
C$_2$S     & 6.9(11) & 2.2(12) & 2.3(12)\\
C$_3$H$_2$ & 8.7(12) & 3.9(13) & 3.6(13)\\
\hline
\end{tabular}
\end{table}
\clearpage
\subsection{Radiative Transfer Model}

The radiative transfer modeling is based on the solution of the radiative
transfer (RT) equation coupled with balance equations for molecular
level populations. We solve this system with the 1D/2D NLTE code
`URAN(IA)' developed by \cite{pavshust}. This code
partly utilizes the scheme originally proposed  and implemented in the
available 1D code `RATRAN' \citep{ratran}.

Here we only summarize the general concept of the `URAN(IA)' code. The
iterative algorithm is the following. First, initial molecular level
populations and a set of photon random paths through the model space
are defined. With these quantities, the specific intensities
$I_{\nu}(i)$ are computed for each cell by the explicit integration of
the RT-equation along the pre-defined photon paths $\vec{n}(i)$. Then,
$\vec{n}(i)$ and $I_{\nu}(i)$ are used to calculate the mean line
intensity $\tilde{J}$ in each cell for all transitions. The computed 
mean intensities are utilized in the next iteration step to refine the
level populations by solving balance equations in all model cells. To
accelerate the convergence of the entire procedure for optically thick
lines, additional internal subiterations for each grid cell (ALI
scheme) are included on top of the global iterations. The adopted
acceleration scheme uses the fact that the calculated mean line
intensity at each particular cell can be divided into an internal component
generated  in the  cell and the external contribution that comes from
other cells of the grid. Subiterations are applied to bring into an
agreement the internal mean intensity of the line and the corresponding
level populations in each cell. After the final molecular level
populations are obtained, we repeat the  calculations, but with
another set of pre-defined random photon paths in order to estimate a
typical error in  the computed values. In our simulations, relative
errors in the level populations are not larger than $1\%$. Finally, the
resulting level populations are used to calculate
excitation temperatures, which are further transformed  into synthetic
beam-convolved single-dish spectra.

\subsection{Evaluation of the Model Quality}

The selection of criteria for a quantitative comparison of modeled and
observed spectral maps represents an important step, but is a rather complicated
problem. In a detailed analysis we would need to take into account a
(dis)agreement of the various features and characteristics of the
spectra, such as intensities, widths, asymmetries, shifts, dips,
regularities in the spatial distributions, etc. These features can be
either analyzed separately or incorporated into a common criterion with
different weights. Using different features and criteria, based on them,
we can assess different aspects of the consistency between the model and observations.

One of the commonly-used techniques to check the model is to compare
distributions of observed and modeled integral line intensities over
the core. This allows to judge how well a model reproduces the
spatial distribution of the total  energy that is emitted in a given
transition, reflecting not only the total molecular content (and excitation
conditions) but also the molecular distributions within the core.

Of course, minimization of this value does not guarantee the consistency
between modeled and observed kinematic properties of the source because
it does not account directly for widths, shifts, asymmetries which
reflect the velocity field. To estimate if the model reproduces the
kinematic structure, e.g., rotation of the core, one can compare the
distributions of the mean velocities. In turn, such a comparison does not
take into account the consistency of the line intensities.

In this paper the quality of the spectra fit is evaluated with the
general criterion
\begin{equation}
{\rm SP}=
\frac{1}{J^{\rm obs}+J^{\rm mod}}\sum\limits_{k=1}^{N_{\rm pos}}
\sum\limits_{i=1}^{N_{\rm chan}}|I_i^{k,\rm obs}-I_i^{k,\rm mod}| \Delta v_i.
\label{ChiSP}
\end{equation}
The inner sum is the absolute difference between observed and modeled spectra at map position $k$,
$I_i^{k,\rm obs}$ and $I_i^{k,\rm mod}$ are
observed and theoretical intensities in velocity channel $i$, $N_{\rm chan}$ is the number of velocity channels, 
$\Delta v_i$ is the channel width, and $J^{\rm obs}$ and $J^{\rm mod}$ 
are the observed and modeled intensities integrated over the frequency and over the map
\begin{equation}
J=\sum\limits_{k=1}^{N_{\rm pos}}\sum\limits_{i=1}^{N_{\rm chan}}I_i^{k}\Delta v_i.
\end{equation}
SP is normalized so that $0\le{\rm SP}\le1$.

\begin{figure}[tbp]
\centering
\includegraphics[width=0.48\textwidth]{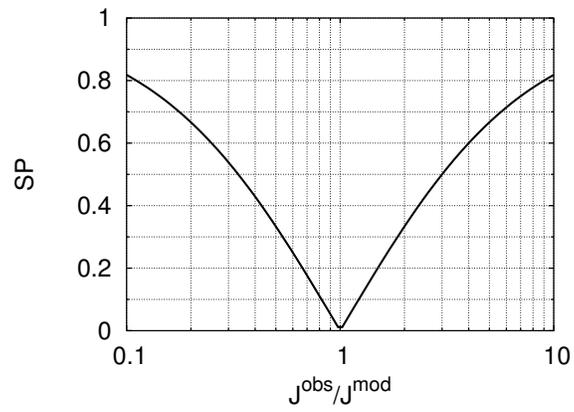}
\caption{
SP criterion as a function of $J^{\rm obs}/J^{\rm mod}$ for
rectangular profiles.}
\label{fly}
\end{figure}

To illustrate the behavior of this criterion, we consider the case of a single spectrum, when
both observed and modeled line profiles are rectangular with equal width $\Delta v$
and position $v_0$, but with different intensities. In
this case SP is equal to
\begin{equation}
{\rm SP}=\frac{|J^{\rm obs}-J^{\rm mod}|}{J^{\rm obs}+J^{\rm mod}}.
\label{baterfly}
\end{equation}
The value of SP as a function of $J^{\rm obs}/J^{\rm mod}$
is shown in Figure~\ref{fly}. In the log-scale this function is
symmetric relative to the point $J^{\rm obs}/J^{\rm mod}=1$ where 
${\rm SP}=0$. If intensities differ by a factor of 2 then 
${\rm SP} \approx 0.4$, while an order of magnitude difference gives
${\rm SP}\approx 0.8$. 

\clearpage
\clearpage


SP, thus, describes the normalized deviation between modeled and observed
line intensities averaged over all velocity channels and all map positions.
On the other hand, SP no longer contains any information about the specific cause
of a disagreement between the model and the observations.

\section{Results for the static model}


Here we consider three important parameters of the static model, namely, the total chemical evolution
time $t_0$, the sticking probability $S$, and the strength of the external UV field $G$.
Regarding these quantities as free parameters we fit the observed spectra, considering the
3D parametric space of models with $0<S<1$, $0<G<1$, and $0.2<t_0<2.0$~Myr, and
calculated 330~models in total.

For each of these models, we calculate molecular abundances as a 
function of radius. Then, distributions of molecular abundances are used as
input data for the RT model.  As a result of the RT simulations, we
obtain distributions of level populations over the core for molecules
of interest, which are further transformed into spectral maps. All
the synthetic line profiles in the map are convolved with
Gaussian beams and shifted in accordance to the observed $V_{\rm LSR}$ position.
Line profiles are calculated for those locations in the map
where they have been observed. Finally, the SP criterion is
checked for each model.

\subsection{Overview of the Core Chemical Structure}

\begin{figure}
\centering
\includegraphics[width=0.8\textwidth,clip=]{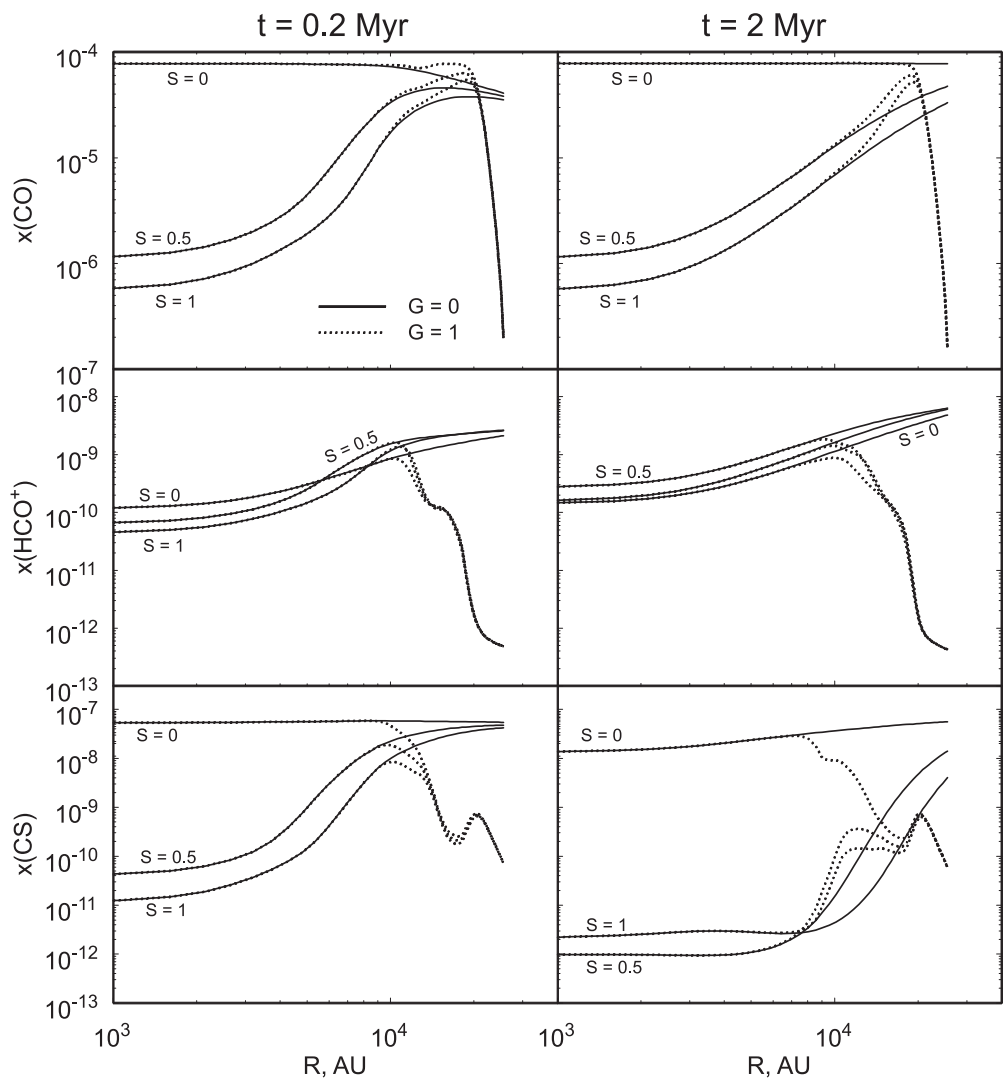}

\caption{Radial distributions of CO (top panels), HCO$^+$ (middle panels),
and CS (bottom panels) for the static model at a chemical evolution time $t=0.2$~Myr (left column) and
2~Myr (right column). Shown are results for no UV and standard UV cases with sticking
probabilities $S=0$, 0.5, and 1.}
\label{stat}
\end{figure}

Radial distributions of CO, CS, and HCO$^+$ abundances for the static
model at $t=0.2$~Myr and $t=2.0$~Myr are shown in Figure~\ref{stat}. The overall chemical structure of the core
is very similar to that obtained by \cite{lee2005}, cf. their Figure~9
(abundances for $t=0$ in their model).
The evolution of the CO abundance is quite simple. In the $S=0$ case (just
gas-phase chemistry) $x$(CO) is almost constant all over the core, being
nearly equal to the total carbon abundance. The non-zero sticking
probability leads to a noticeable CO freeze-out in the core and to a
minor CO depletion in the envelope. In the case with UV
illumination CO molecules in the envelope are almost totally
dissociated.

The HCO$^+$ abundance is not very sensitive to the actual $S$ value.
The only parameter that can be more or less reliably
constrained with observations of HCO$^+$ is $G$.

The CS abundance depends on all three
parameters. The location of the dip in the radial CS profile,
appearing in UV-illuminated models, coincides with the region of
enhanced CO abundances, which decrease the number of carbon atoms
available for CS formation.

\clearpage
\clearpage

The discussed behaviour is not strongly
time-dependent, being only slightly more pronounced at later times of the chemical evolution.
We should note that the same general features are shared by dynamical models,
which justifies our usage of line profiles in the static model as
a guide for possible ranges of $S$ and $G$.

\subsubsection{Note on N-bearing Species}

Before we analyze the CO, CS, and HCO$^+$ data, we discuss briefly
our ability to match the other available observations, specifically, the data on N-bearing
species that are not the main subject of the current study. The column density of N$_2$H$^+$
has been determined by \cite{benson1998} and \cite{cas2002} to be $3-5\times10^{12}$~cm$^{-2}$.
The column density of NH$_3$ is about $8\times10^{14}$~cm$^{-2}$ \citep{lemme1996,jijina}.
Our values, compiled in Table~\ref{comparaikawa}, are somewhat higher for both molecules. Currently,
our ability to treat these molecules with URAN(IA) is limited because of the poor knowledge
of hyperfine transition parameters. We
modeled the N$_2$H$^+$(1-0) profile, neglecting its hyperfine structure and obtained antenna temperatures
of the order of 0.4--0.6~K, similar to what is observed. We also performed an approximate
modeling of the HCN (1--0) line and compared the result with the profile obtained by \cite{TurnerVIII}.
We are able to reproduce the antenna temperature and main features of the profile, in
the sense that both the $F=2-1$ and $F=1-1$ transitions are self-absorbed, while the $F=1-0$
transition is not self-absorbed. The intensity ratios in our model are different from those in
\cite{TurnerVIII}. Molecular data from \cite{moldata} are used for this analysis.

\subsection{Optically Thin Lines}

\begin{figure}
\centering
\includegraphics[width=1.0\textwidth]{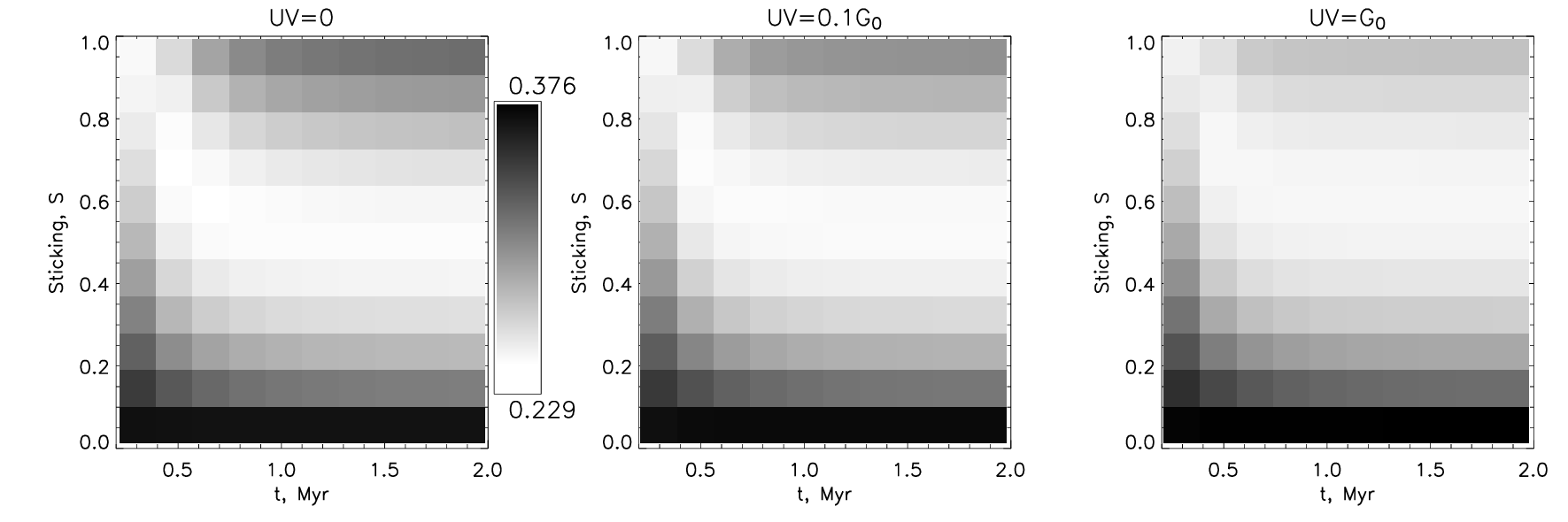}
\includegraphics[width=1.0\textwidth]{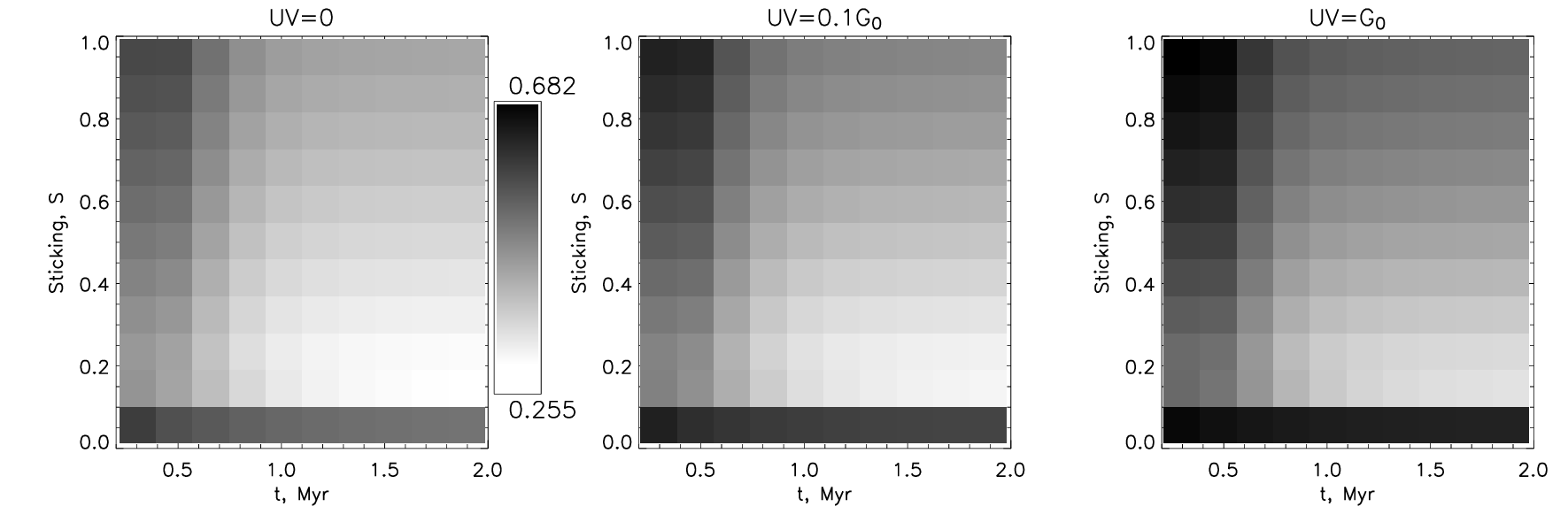}
\includegraphics[width=1.0\textwidth]{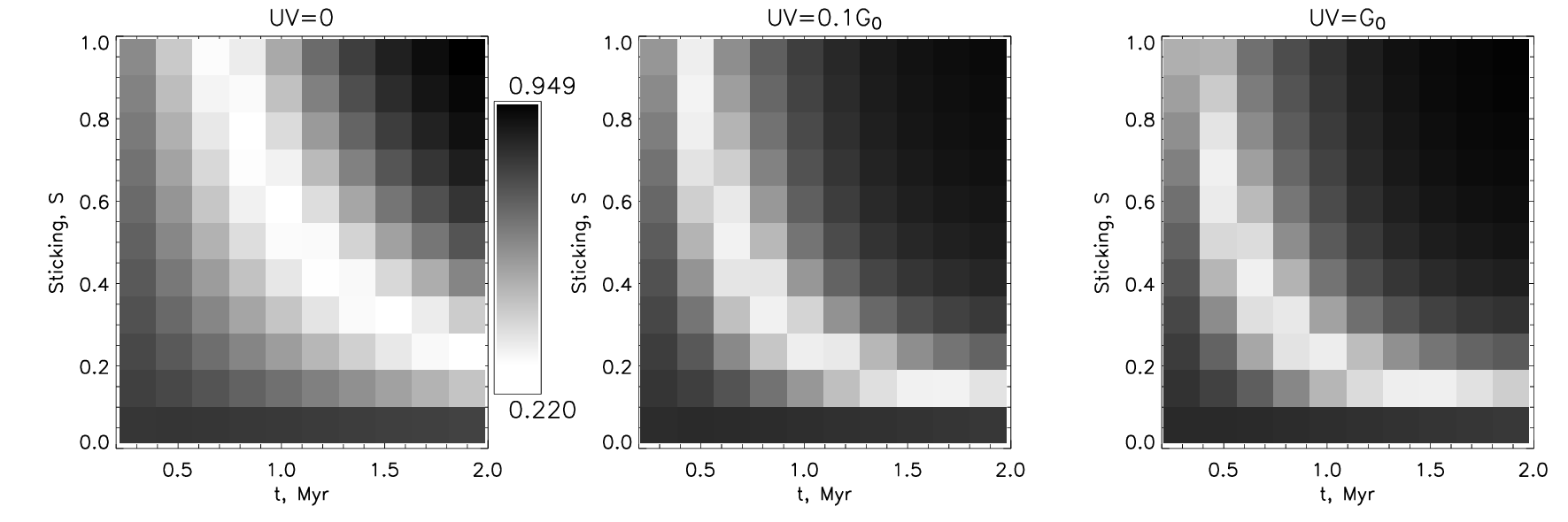}
\caption{Values of SP-criterion ($ts$-diagrams) for optically thin transitions of C$^{18}$O (top row),
H$^{13}$CO$^+$ (middle row), and C$^{34}$S (bottom row). Smaller values 
(lighter colors) correspond to better agreement. Columns
differ by $G$ values, indicated on top of each panel.  The gradient
scales on the  right side of the first  plot in each row are appropriate
for all plots in the row.}
\label{chispfig}
\end{figure}

We first analyze optically thin transitions that probe the total
molecular content of the core. In order to show the fitting results
in a compact way, we present them as $ts$-diagrams which are
2D-plots of SP-values for various $t_0$ and $S$ parameters at a fixed
$G$ value. The $ts$-diagrams for optically thin
transitions of C$^{18}$O, H$^{13}$CO$^+$, and C$^{34}$S are shown in
Figure~\ref{chispfig}.

\clearpage
\clearpage

All panels in Figure~\ref{chispfig} indicate that values of $S$ close to zero are
ruled out. Higher sticking efficiencies are not favored in the analysis
of the C$^{18}$O and H$^{13}$CO$^+$ transitions. However, it must be kept in
mind that the range of SP is not very large for these
transitions. This means that conclusions from their analysis must be taken with
care. 

The C$^{34}$S lines appear to be more robust discriminators between the
model parameters. On the $ts$-diagram we clearly see the hyperbolic zone
of best fitting with a significant range of $0.22<{\rm SP}<0.95$.
As might have been expected, smaller $S$ values become
appropriate as $t_0$ increases, as a lower sticking rate is compensated
by a longer timescale.

In general, there are combinations of the studied parameters that
provide a good agreement with observations, with SP values being
less than 0.3 for all the optically thin transitions simultaneously.
This corresponds to differences in the line intensities by a factor of~2.
All the considered optically thin lines are not very sensitive to
variations of the UV field. Indeed, they are mostly formed in the 
inner part of the core where the UV field is attenuated by the envelope.

\subsection{Optically Thick Lines}

Although the static model is quite successful in reproducing the observed optically
thin line profiles, this should not be overinterpreted.
Both the observed and modeled
optically thin lines are nearly Gaussian in shape. The width of the
theoretical profiles is defined by the adopted value of the
microturbulent velocity. In our study this velocity is chosen to be 0.15 km s$^{-1}$
in order to fit the observed line widths. Thus, for optically thin
lines the quality of the fit in a static model depends mainly on the
column density.

\begin{figure}
\centering
\includegraphics[width=1.0\textwidth]{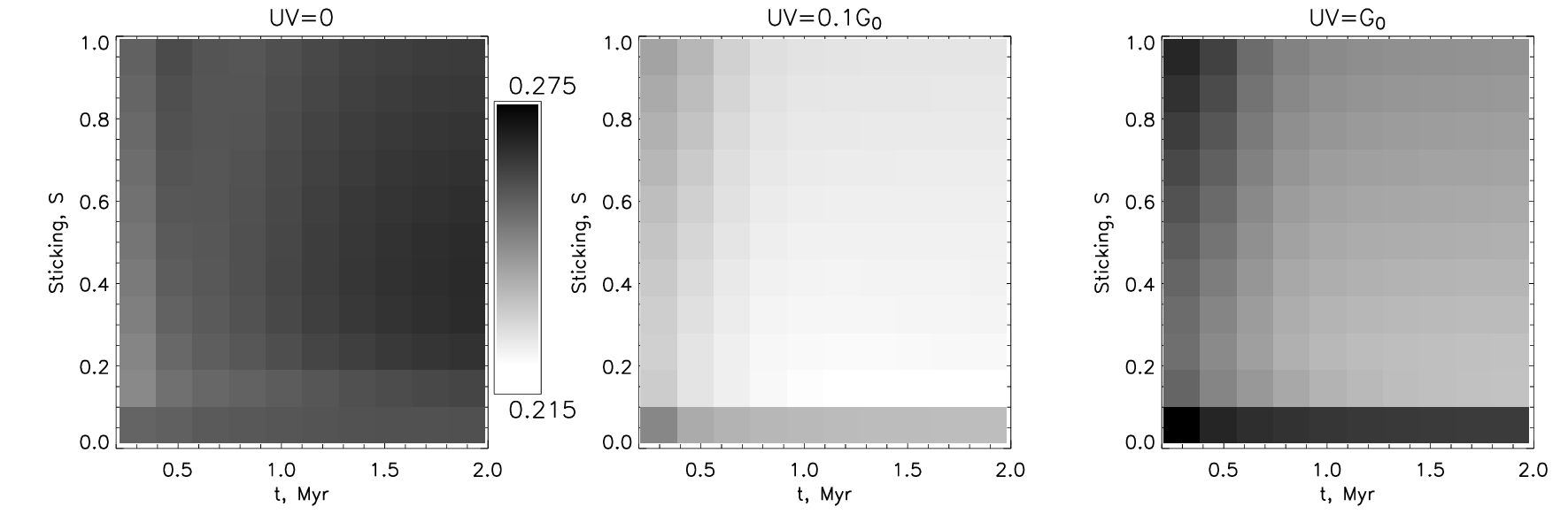}
\includegraphics[width=1.0\textwidth]{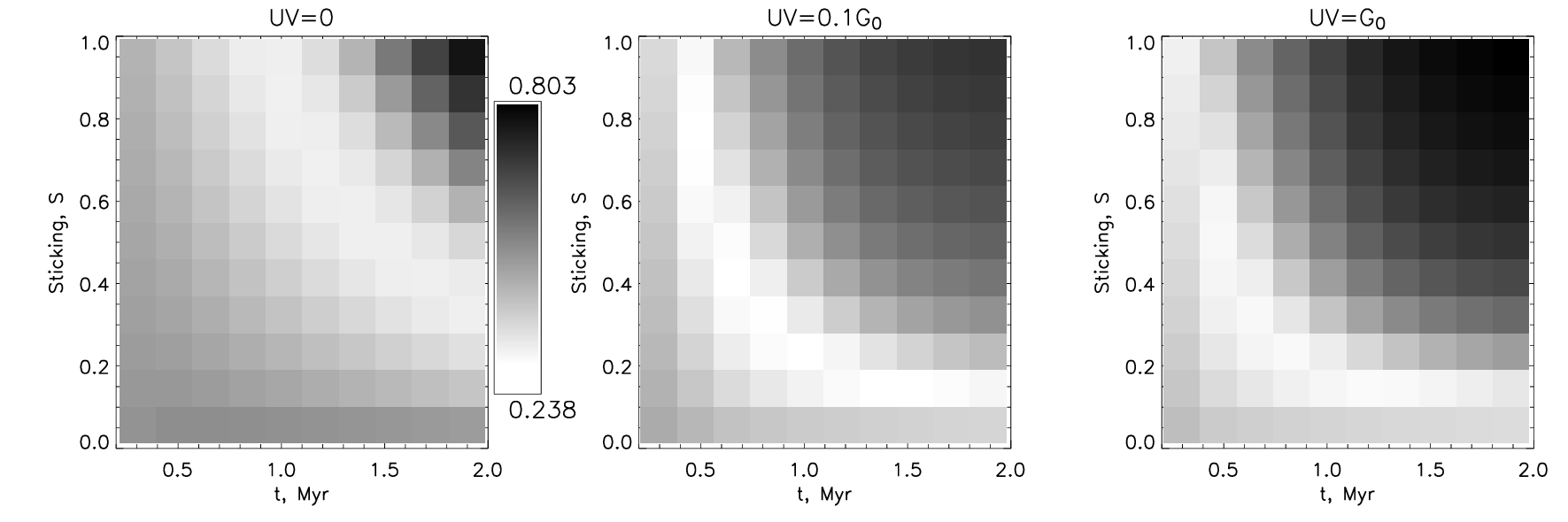}
\caption{Values of SP for the optically thick transitions of HCO$^+$ (top row) and CS
(bottom row). Smaller values (lighter squares) correspond to
better agreement. Columns differ by $G$ values indicated on the top of
each panel.}
\label{phickfig}
\end{figure}

On the other hand, optically thick transitions are mostly sensitive to
the conditions in the envelope, which makes them promising
tracers of the external UV field. In our study, we consider the
optically thick transitions of CS and HCO$^+$. The $ts$-diagrams for these
transitions are shown in Figure~\ref{phickfig}.

\clearpage
\clearpage

As expected, both the HCO$^+$ and CS diagrams are sensitive to variations of
the UV radiation. In the case of no UV field (left panel), molecules survive in the 
envelope and produce emission over a broad range of impact parameters.
In addition, enhancement of the molecules in the envelope leads to
prominent dips in the profiles of optically thick
lines. In contrast, a strong UV field, $G=1$ (right panel), destroys
molecules in the outer parts of the core, leading to more centrally peaked
distributions of the integral intensity and to weaker self-absorption
dips. Both effects worsen the agreement between observed and synthetic
maps.

\begin{figure}
\centering
\includegraphics[width=0.6\textwidth]{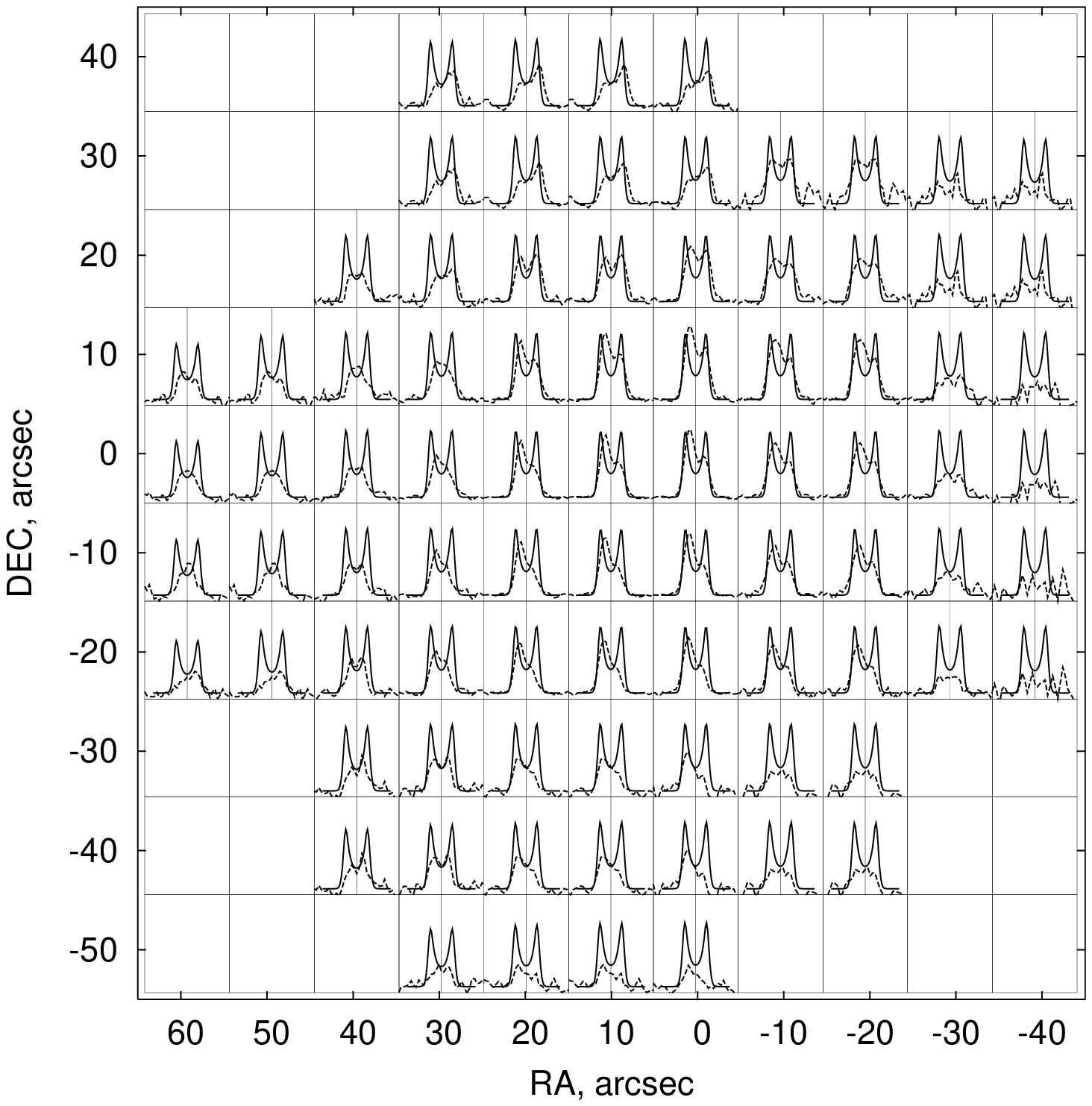}
\includegraphics[width=0.6\textwidth]{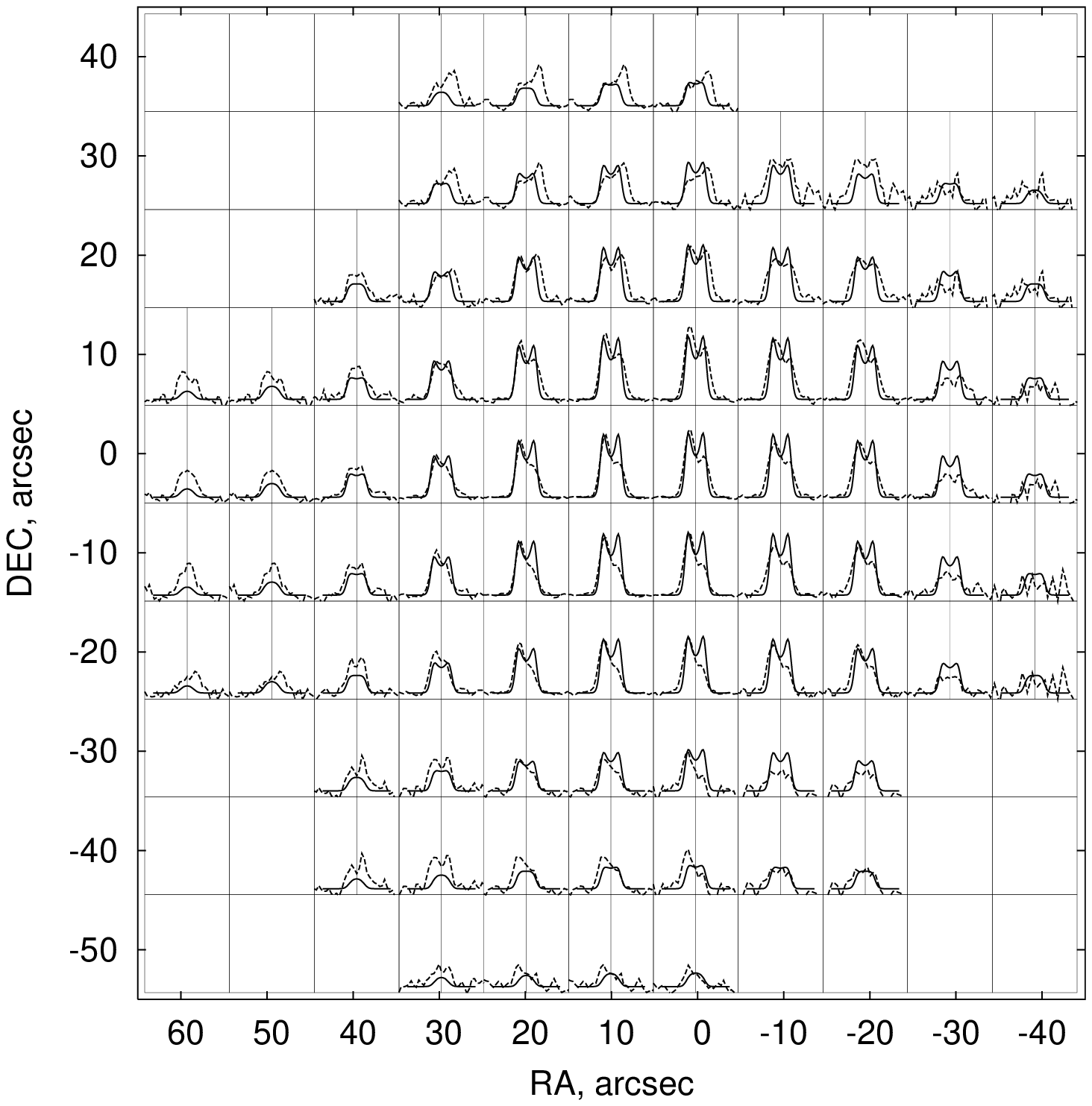}

\caption{The map of the observed (blue) and modeled (red) CS(2-1) line profiles. 
In the top panel lines are calculated for the static model with 
$G=0$, $S=0.6$  $t=0.8$~Myr. The bottom panel corresponds to the static model with 
$G=1$, $S=0.6$  $t=0.4$~Myr.}
\label{stat3}
\end{figure}

To demonstrate the influence of the UV field on the spectra,
we show observed and theoretical spectral maps of CS(2-1) in
Figure~\ref{stat3}. For
the static model with no UV radiation, $S=0.6$ and $t=0.8$~Myr
CS line profiles have nearly the same intensity at all positions
and self-absorption dips which are much deeper than
in the observed spectra. The synthetic intensities of the 
model with strong UV field, $S=0.6$, and $t=0.4$~Myr decrease rapidly toward the edge of the
core. The self-absorption dips become shallower than in the
model without UV field.

\clearpage
\clearpage

According to Figure~\ref{phickfig}, the model with attenuated
UV radiation  ($G=0.1$) seems to be most appropriate. It must be noted
that it would be impossible to constrain the effect of the UV field using a central spectrum
alone. One may argue that the excess intensity of the CS(2--1) line in the zero UV model
can be mediated by an adjustment of the sulfur abundance which
is not well constrained observationally. We computed two additional series of models
with zero UV intensity and sulfur abundance decreased by a factor of 3 and 10,
relative to our standard value. These models showed that a lower initial sulfur
abundance leads to an overall decrease of S-bearing species in the core. These
results in a better agreement for the central spectrum. However, the gradual
decrease of the CS line intensity toward the outer boundary of the core can only be reproduced in models
with some UV illumination.

Although the modeled optically thick lines have the expected double-peaked shape, these peaks are
equal in height, contradicting the observed asymmetry. In order to
describe this asymmetry, we must move from the static model toward a
dynamical model. From Figures~\ref{chispfig} and \ref{phickfig} it is obvious that
the sticking efficiency is not well constrained in the static model. Therefore, the only parameter
we will fix in dynamical models is the strength of the UV field ($G=0.1$).

We will also not consider $t_0<0.4$~Myr values as they do not fit both the optically thin
and optically thick lines. This seemingly minor limitation has an important implication
for our dynamical models. The asymmetry of the central spectrum mentioned in the
previous paragraph implies that the CB17 core undergoes an infall with a velocity of
$\sim0.05$--0.1~km s$^{-1}$. In the adopted prescription and for $t_0>0.4$~Myr such velocities
are only possible with $\delta\ge0.5$.

\section{Results for the dynamical model}

As described in \S~\ref{dynmodsect}, the dynamical history of the core is
represented by two parameters, which are the evolutionary time $t_0$ and the
power law index $\delta$. In the following, we investigate the sensitivity
of molecular spectral maps to these parameters, at the same time
searching for combinations of $t_0$ and $\delta$ which give the
best agreement between observed and modeled CB17 maps.

Ideally, in order to find the `best-fit' parameters for the CB17 core
one should vary $G$, $S$, $t_0$ and $\delta$, simultaneously. However,
the primary goal of this paper is rather to show the general effect of the
selected parameters.
Therefore, we fix $G$ to a value of 0.1 as determined from the static models, but vary $S$ together
with the dynamical parameters $t_0$ and $\delta$.

In the dynamical modeling yet another parameter has to be considered.
In the static configuration we assumed the microturbulent velocity $V_{\rm turb}$ to be 0.15 km s$^{-1}$.
This value is needed to reproduce the line width in the static model and
as such it hardly leaves any room for a systematic velocity field.
On the other hand, the central line asymmetry clearly shows that
there is a non-zero infall velocity in the CB17 core. This seems
to imply that $V_{\rm turb}$ is to be varied along with the other
parameters. On the other hand, it obviously makes no sense to consider
those values of $V_{\rm turb}$ which in combination with the systematic
velocity would result in line widths that are too large or too small.

To minimize the needless effort, we adopted the following approach. For
each combination of $\delta$ and $t_0$ we select $V_{\rm turb}$ to get
the line width of about 0.15~km s$^{-1}$, so that the differences between
theory and observations are only caused by disagreements in the line
intensity and in the relative heights of blue and red peaks in optically
thick lines.

\subsection{Collapsing Core without Rotation}

\begin{figure}
\centering
\includegraphics[width=\textwidth]{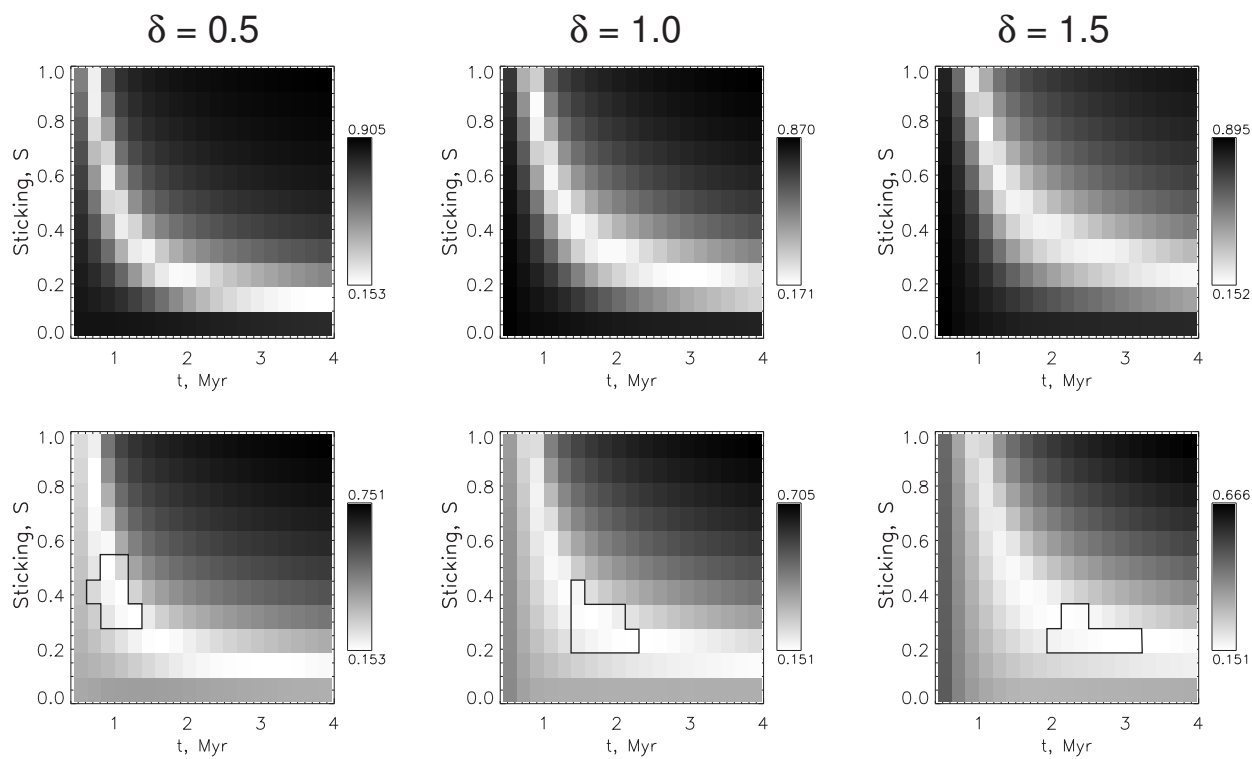}
\caption{Values of SP for the optically thin C$^{34}$S transition (top row) and
the optically thick CS transition (bottom row). Columns differ by $\delta$ values indicated on top of
each panel. `Best-fit' regions, selected by the blue-to-red
peak intensity ratio, are shown by solid contours in the bottom row.}
\label{norot}
\end{figure}

In purely collapsing models
we only compare our results to the central spectrum of CB17 to avoid
confusion with the effects of rotation. Results of the modeling
are presented in Figure~\ref{norot}. We show
SP values for three values of $\delta$ (0.5, 1.0, 1.5), vary $t_0$ between
0.4 and 4~Myr and $S$ between 0 and 1. Only results for CS lines
are given, as other molecules demonstrate much less sensitivity for
the discussed parameters.

\clearpage
\clearpage
Figure~\ref{norot} does not show any significant
differences to the static model (see Figure~\ref{chispfig}, bottom panel,
and Figure~\ref{phickfig}, bottom panel). In models with increasing $\delta$ the
`best-fit' region shifts toward later times which is expected,
as larger $\delta$ values lengthen the `less-dense' stage of core
evolution. It seems that we have the same dilemma as in the static case, being
unable to distinguish between large $S$ and large $t_0$. However, if we look
at numeric SP values within the hyperbolic area, we see that the best-fit model
for the CS transition (which is most sensitive to the infall velocity) corresponds
to $S=0.2-0.3$ for all three $\delta$ values.

The theoretical line profiles show that the agreement
is still not perfect even in the `best-fit' region. In some cases
the line widths are different, in other cases
the ratio between blue and red peaks is not the same as observed.
The line width can be further adjusted with a more appropriate
choice of $V_{\rm turb}$. Thus, to split the contribution of the line width and the peak asymmetry in a
`residual' disagreement, we consider an additional criterion which is just the difference
between the observed and the theoretical blue-to-red peak intensity ratios. Contours
in the lower part of Figure~\ref{norot} indicate models with the best agreement
between theory and observations both in terms of the overall profile shape and
the profile asymmetry. The combined selection procedure leads to the model parameters
summarized in Table~\ref{tabdynpar}. Minimum SP values are selected only among locations within
contours. The best-fit models are somewhat different for CS and C$^{34}$S.
\clearpage
\begin{table}
\caption{The `best-fit' parameters for the dynamical model without rotation.}
\label{tabdynpar}
\begin{tabular}{lllll}
\hline
$\delta$ & $t_0$, Myr  & $S$ & SP(CS) & SP(C$^{34}$S) \\
\hline
0.5      & 1.0--1.2         & 0.3--0.4 & 0.153  & 0.208         \\
1.0      & 1.6--1.8         & 0.3 & 0.152  & 0.187         \\
1.5      & 2.2--2.4         & 0.3 & 0.154  & 0.190         \\
\hline
\end{tabular}
\end{table}
\clearpage
The data in Table~\ref{tabdynpar} indicate that the sticking probability
has an effective value of about 0.3, while the age of the core is larger than 1~Myr and probably less than 2.5~Myr.
All presented SP values are smaller than in the corresponding
static model. However, it must be kept in mind that these numbers are not directly
comparable as they are computed for the central spectrum only
in the case of the dynamical model, while in the
static model SP values are calculated for the whole map.

The difference between SP values computed for various $\delta$ is small for
C$^{34}$S and only marginal for CS. However, both transitions show a minor
preference toward the model with $\delta=1.0$. In the next subsection this model
is used as an input for the 2D model. We emphasize that in the general case one may
have to vary $V_{\rm turb}$ along with the other parameters.

\subsection{Collapsing Core with Rotation}

Finally, we now include rotation in the dynamical
description of the core, the RT problem becoming two  dimensional. In our model
(\S~\ref{rotomega}), the rotation
of the core is characterized by the initial angular
velocity $\Omega$. To simulate observations, we  also have to specify
the orientation of the core with respect to the observer, which is
defined by the inclination angle $i$  and the position angle PA. We assume that the
core is observed edge-on ($i=90^{\circ}$) in order to get an estimate of the minimum
value of its angular momentum. No attempt
is being made to reproduce the inclination angle as within the framework of a spherically
symmetric model we are not able to distinguish between rapid rotation and low inclination.
We tried several other values for $i$ and found that variations of excitation conditions
due to different rotation velocity profiles in the CB17 model are not strong enough
to produce noticable differences in observed spectra. It would be possible to determine $i$,
using a more realistic dynamical model which would provide an independent information
on the core rotation. Also, we note that it may be possible to estimate $i$ independently for elongated
cores from geometric reasoning.

The two parameters that are 
varied to get the best agreement with observations are the initial
angular velocity $\Omega$ and the
position  angle PA. The position angle is defined as the angle between the projection of
the core rotation axis on the sky and the direction to the North.

The simulation is made in the following way. The chemical structure of
the core and its radial velocity field are taken from the best-fit 1D
dynamical model with the parameters $\delta=1.0$ and $t_0=1.6$~Myr. For
the given $\Omega$ value we generate azimuthal velocities according to
Eq.~(\ref{Vphi}). These  velocities are combined with the infall
velocities  to get the 2D velocity field. This field is then used to
solve the 2D radiation transfer equation. Having specified the PA, we 
calculate the convolved spectral map, which is further compared to the observed
map using the SP criterion.
\clearpage
\begin{figure}
\includegraphics[width=0.4\textwidth]{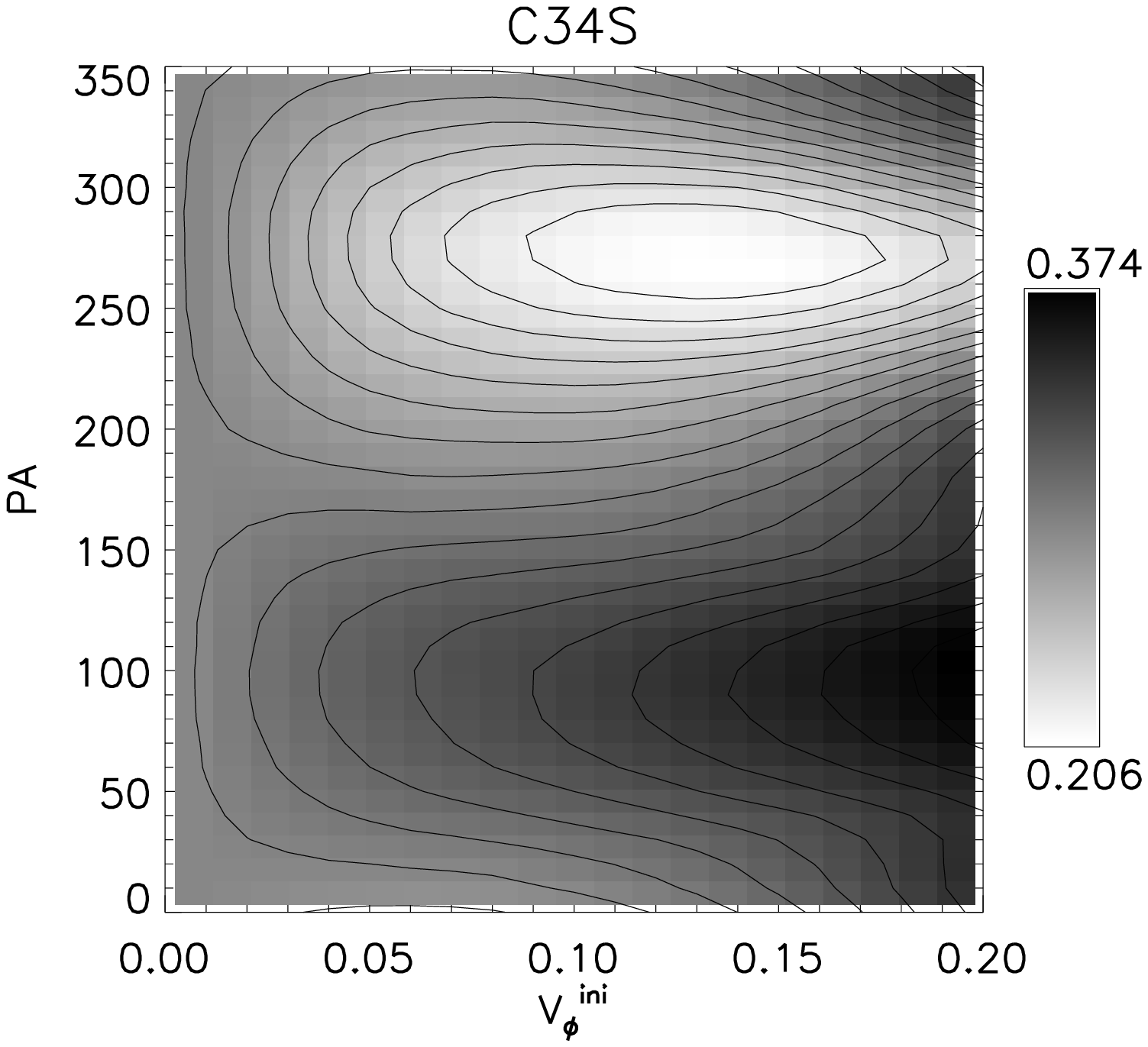}
\includegraphics[width=0.4\textwidth]{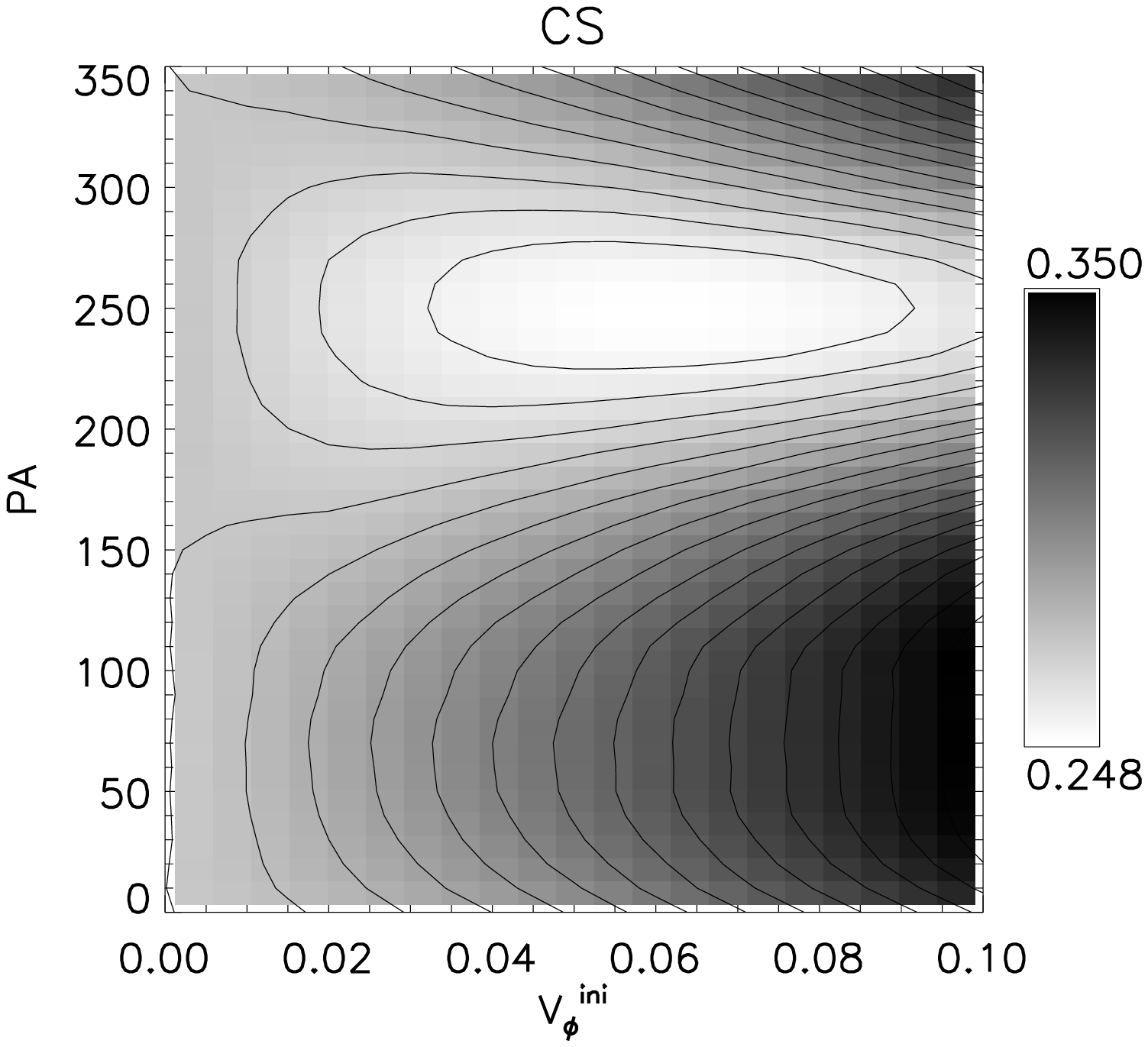}
\caption{Values of SP for the optically thin C$^{34}$S transition 
(left) and for the optically thick CS transition (right) for
various values of $\Omega$ and PA. Smaller  values (lighter
squares) correspond to better agreement. The $x$-axis is labeled in units
of the initial linear rotation velocity at the edge of the core. A velocity
of 0.1~km s$^{-1}$ corresponds to $\Omega=2.7\times10^{-14}$~s$^{-1}$.}
\label{paomega}
\end{figure}
\clearpage
Results of this comparison are shown in Figure~\ref{paomega}. Obviously,
when there is no rotation, the model shows no sensitivity to the PA value.
As we increase $\Omega$, both the optically thin and optically thick
transitions show a clear preference to a PA value of $250^{\circ}-300^{\circ}$.
It is interesting to note that the sensitivity of the model to the position
angle increases when we adopt $\Omega$ values that are greater than
allowed observationally.

When the position angle is specified incorrectly, both transitions
indicate that the model with pure infall agrees better with the observations
than the model  with infall and rotation (SP values increase from left
to right in Figure~\ref{paomega}). For ${\rm PA}\approx270^{\circ}$, SP
values have minima at $V_{\phi}^{\rm ini}=0.13$ km s$^{-1}$ for the C$^{34}$S
line and at a somewhat
smaller value of $V_{\phi}^{\rm ini}=0.07$ km s$^{-1}$ for the CS
line. The disagreement is probably caused by a higher rotation
velocity closer to the core center  (where the C$^{34}$S line is generated)
than our model predicts. This can be further adjusted by adopting a
different initial rotation velocity distribution. However, it must be
noted that the  optically thick CS line is in general less sensitive to
$\Omega$ variations which is indicated by the smaller SP range. This
implies that optically thin lines are better rotation  indicators, at
least, when the SP criterion or another similar width-sensitive
criterion is used.

We note that the lowest SP values in  the model
with infall and rotation are actually higher than the corresponding
values for the  static model. This is caused by the values for $V_{\rm turb}$
adopted in the dynamical model.  Because we did not try to vary
$V_{\rm turb}$ along with the other parameters, the widths of some 
theoretical profiles do not quite fit those of their observed
counterparts. This results in higher SP values.
\clearpage
\begin{figure}
\includegraphics[width=0.4\textwidth]{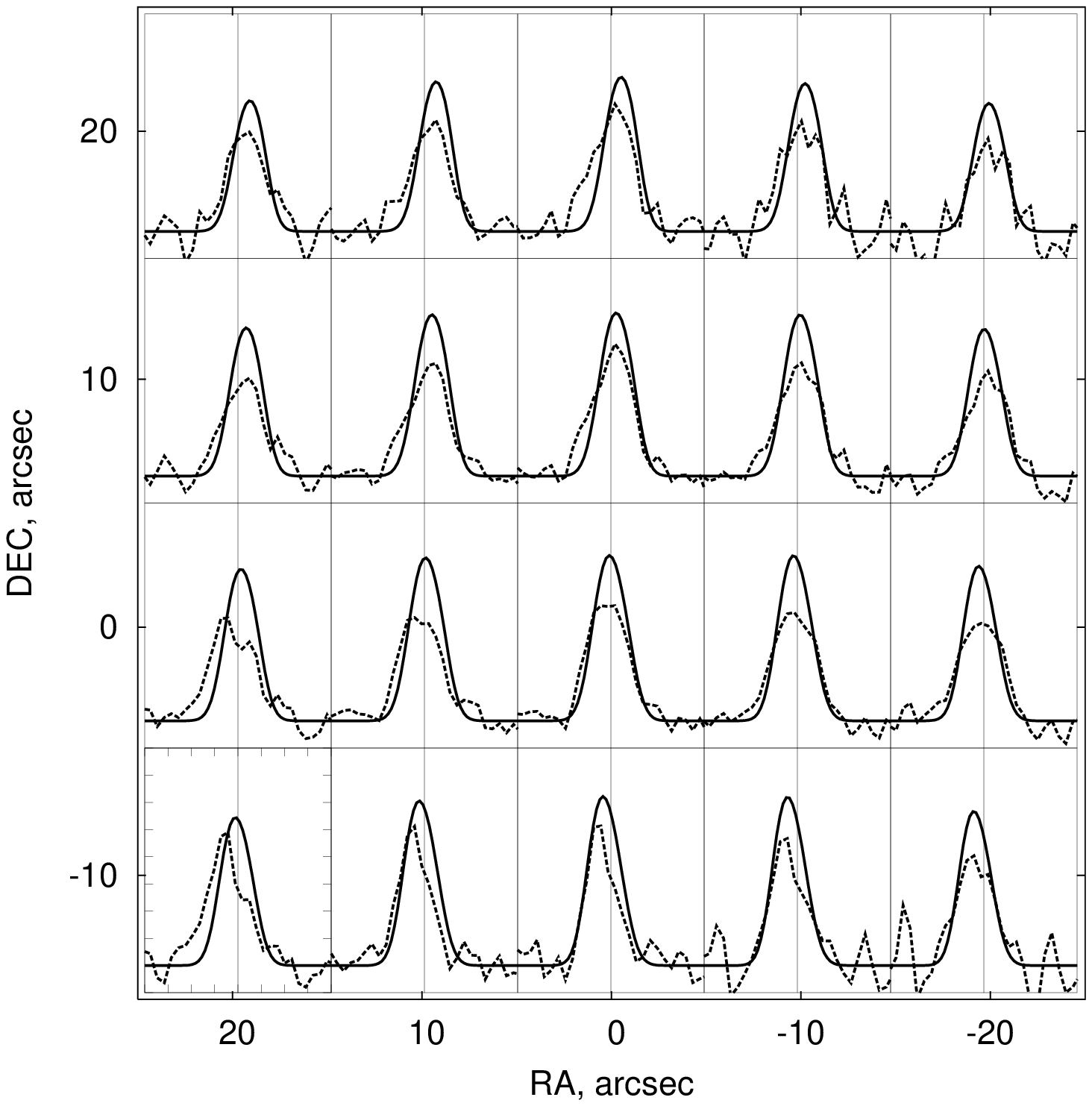}

\includegraphics[width=0.4\textwidth]{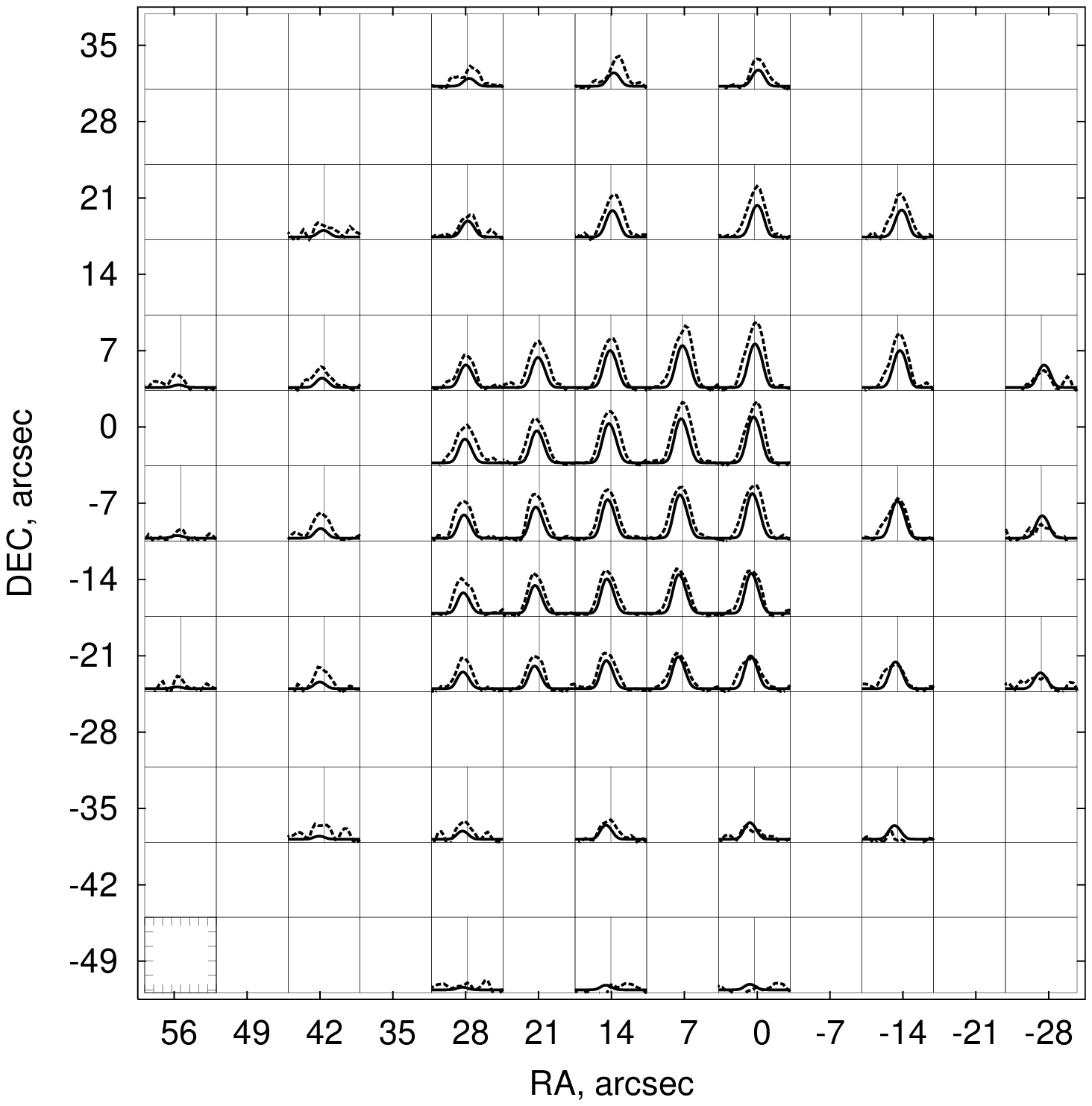}

\includegraphics[width=0.4\textwidth]{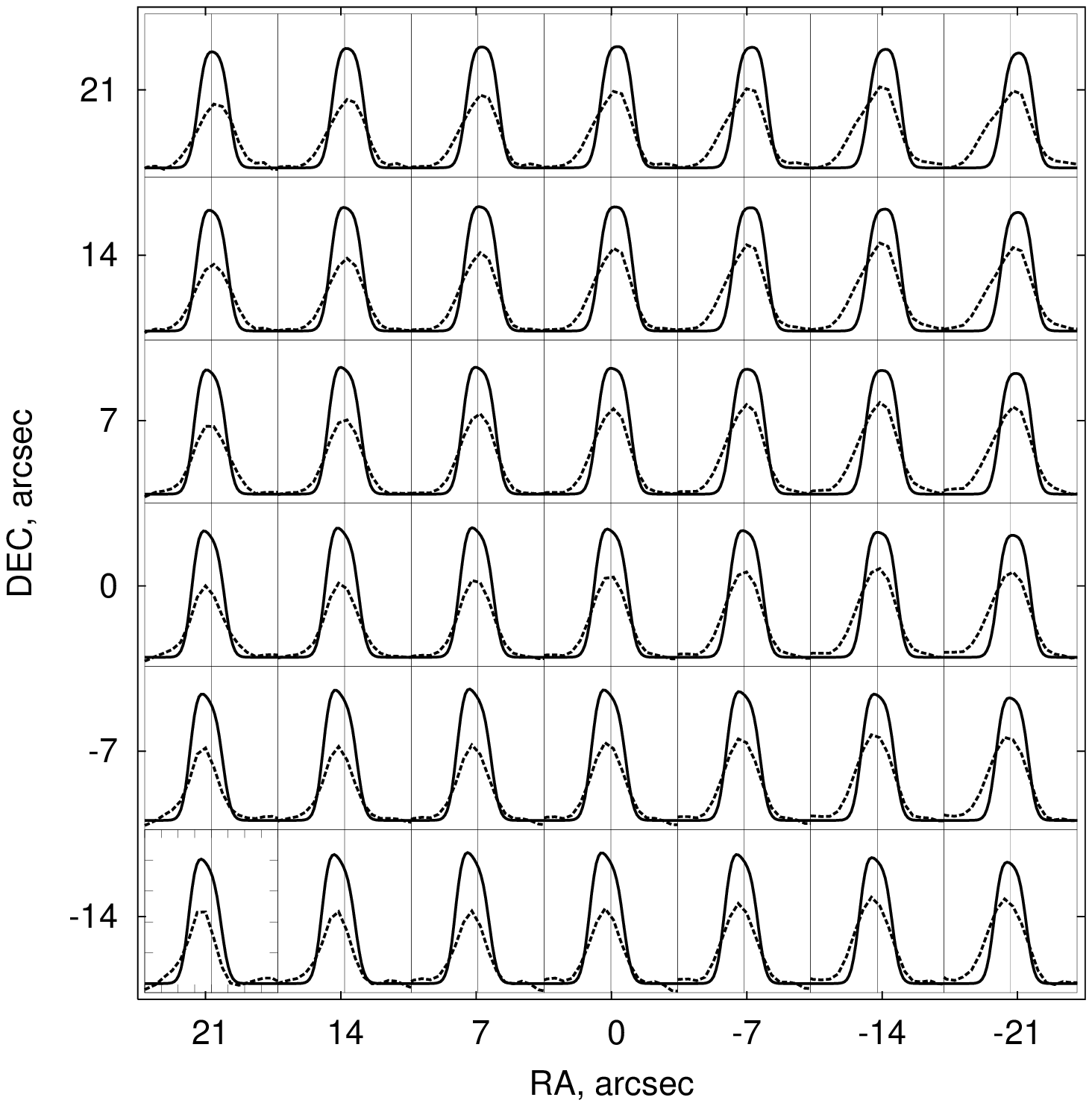}

\caption{Spectral maps for the `best-fit' CB17 model. Shown are optically thin transitions
(from top to bottom---C$^{34}$S, H$^{13}$CO$^+$, C$^{18}$O).}
\label{bestmaps1}
\end{figure}

\begin{figure}
\includegraphics[width=0.6\textwidth]{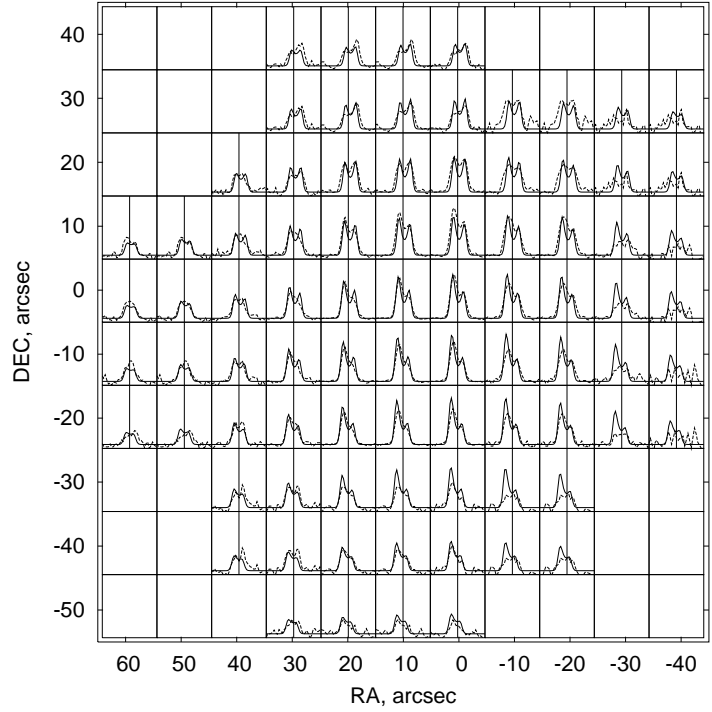}

\includegraphics[width=0.6\textwidth]{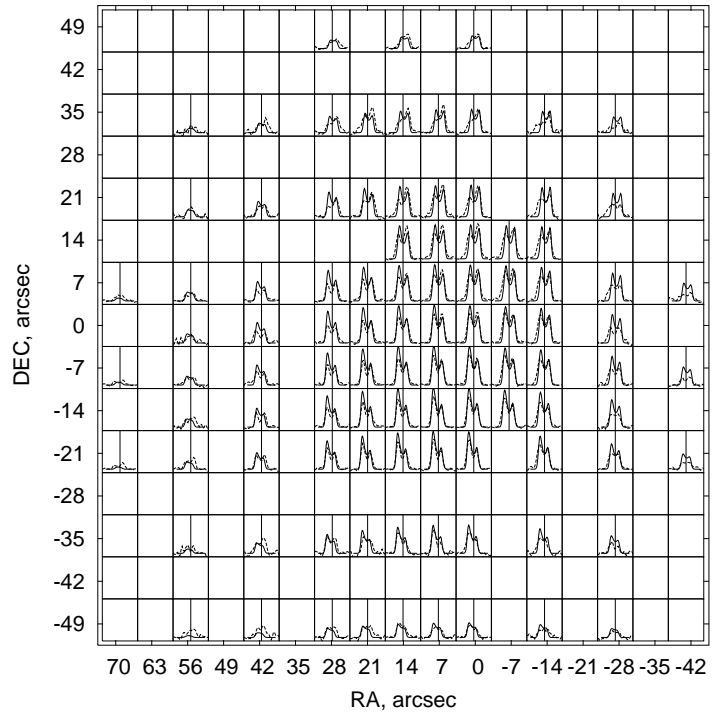}

\caption{Same as in Figure~\ref{bestmaps1} but for
optically thick transitions (from top to bottom---CS, HCO$^+$).}
\label{bestmaps2}
\end{figure}
\clearpage
Finally, in Figures~\ref{bestmaps1} and \ref{bestmaps2} we show the convolved spectral maps of the
CS(2-1), C$^{34}$S(2-1), HCO$^+$(1-0), H$^{13}$CO$^+$(1-0), and C$^{18}$O(2-1) lines
for the `best-fit' 2D model (Table~\ref{tabdynrotpar}).
In all the modeled optically thin transitions shifts of the observed lines are reproduced
quite successfully. Furthermore, the modeled optically thick CS profiles
follow the asymmetry of the observed spectra, with blue-shifted profiles in the 
left bottom corner and nearly equal peak intensities in 
the upper part of the map. The consistency between the
shapes of the theoretical and observed HCO$^+$ profiles is less
impressive. However, the theoretical line intensities are close to the observed ones,
and the asymmetry is well reproduced in the central part of the map.
The fit is worse in the upper part of the map, where there is some disagreement
in the CS profiles as well. Given the overall consistency between maps of optically thin
and optically thick lines, we believe this residual disagreement originates in
the core structure which is certainly more complicated than the assumed spherical
symmetry. It is also possible that there are some small-scale motions within the
core that are not well described by the microturbulence model (like the motions
discussed for B68 and L1489, see Introduction).
\clearpage
\begin{table}
\caption{The best-fit parameters for the dynamical model with rotation.}
\label{tabdynrotpar}
\begin{tabular}{ll}
\hline
Parameter & Value \\
\hline
Chemical age & $\sim2$  Myr \\
The UV-field ($G$) & 0.1 \\
Effective sticking probability & 0.3 \\
$\delta$ & 1.0 \\
$V_{\rm rot}$ & 0.1 km s$^{-1}$\\
$V_{\rm infall}$ & 0.05 km s$^{-1}$\\
$V_{\rm turn}$ & 0.1 km s$^{-1}$\\
PA & $250^{\circ}$ \\
\hline
\end{tabular}
\end{table}
\clearpage
In general, all the modeled maps do reproduce the spatial distributions of 
line intensities, except the C$^{18}$O(2-1) intensity, which is approximately
1.5--2 times greater than that observed. The C$^{18}$O overabundance is
present in the static model as well and seems to follow from the adopted
relatively low $S$ value which is supposed to provide the best agreement
for CS lines. Upper panels in Figure~\ref{chispfig} indicate that
regardless of UV intensity values $S>0.5$ are preferred from the point of
view of the C$^{18}$O abundance.

Some further adjustment of the CO abundance can be made with
a variation in the cosmic ray ionization rate.
However, there is a more natural explanation. In our study we assume that the
C$^{16}$O/C$^{18}$O ratio is constant and equal to 490 which is just the $^{16}$O/$^{18}$O
ratio. Due to the chemical fractionation
the ratio C$^{16}$O/C$^{18}$O can vary by a factor of a few \citep{federman}.
Specifically, the isotope-selective photodestruction of CO
molecules may decrease the abundance of the isotopomer with the less abundant isotope
\citep{bl1982}. Importantly, the fractionation apparently affects the C$^{16}$O/C$^{18}$O
ratio stronger than the $^{12}$CO/$^{13}$CO ratio \citep{chuwatson,federman}, which would explain
why we have the intensity disagreement in the C$^{18}$O lines, but not in
the H$^{13}$CO$^+$ lines.

\section{Discussion}

This paper represents an attempt to develop a technique  that allows to
get as much information as possible from the spectral maps of a starless
core. The technique is outlined in Figure~\ref{flchart}. As an example,  we used the CB17 core for which we have detailed
observational data. This core  appears to be quite young, having a
chemical age of about 2~Myr. It is  tempting to put it into an
evolutionary sequence with other starless cores. Taking the
central density as an indicator  of the dynamical evolution, we have to
assume that CB17 is quite  evolved as it is almost as dense as L1544. 
From the point of view of the central depletion CB17  appears to be less
evolved. There is some depletion in its center, which  is indicated by
high SP values for models with $S=0$. In  best-fit static models (e.g.,
$t_0=1$~Myr, $S=0.4$) the CO molecule is depleted by  a factor of 40 in
the core center, and the CS molecule is depleted by a similar  factor.
However, this depletion is not very obvious in the CS column 
density distribution and is not seen at all in the CO column density 
distribution. This explains why the central holes are not seen in the 
observed maps of optically thin transitions of
C$^{34}$S and C$^{18}$O. Within the classification 
scheme suggested by \cite{lee2003} and \cite{shirley}, the CB17 core is
{\em  evolved dynamically\/} and {\em moderately evolved chemically}.

Within the  framework of the adopted dynamical formalism we notice that ages 
greater than $\sim0.5$~Myr and infall speeds of the order of 0.05~km 
s$^{-1}$ are required. This implies $\delta>0.5$ (the kinematic difference 
between the considered $\delta$ values of 0.5, 1.0, and 1.5 is less 
significant). In this collapse regime the central density first stays almost 
constant and then undergoes  a stage of a very fast growth. This conclusion 
is further illustrated by the comparison of the derived chemical age, 
$t_{\rm chem} \approx 2$~Myrs, to the CB17 free-fall timescale, $t_{\rm 
ff}$. For the assumed initial density of $n($H$_2)=5\cdot 
10^3$~cm$^{-3}$, $t_{\rm ff}=0.4$~Myrs which is five times smaller than 
$t_{\rm chem}$. So, formally we can conclude that the core has evolved more 
or less quasi-statically for $\sim10^6$ years, then quickly lost stability 
and started to collapse. It seems reasonable just from mass conservation 
considerations  that this is the appropriate scenario for a dense, 
collapsing, and chemically mature core. The scenario is similar to that favored by 
\cite{shem} and considered by \cite{lee2004}. The quasi-static evolution of 
the starless core can result from the magnetic field support being 
gradually lost due to the ambipolar diffusion. On the other hand, \cite{aikawa2005}
 and \cite{KetoField2005} showed that quasi-static evolution of 
prestellar cores can be explained by the assumption that the initial condition for contraction
represent a destabilized Bonnor-Ebert sphere.

The estimated CB17 age of 2~Myr may have implications for current theories 
of the star formation. For example, it is greater than the typical lifetime 
of dense cores found in gravoturbulent simulations \citep{vslifetimes} and 
the observationally estimated lifetimes of submillimeter cores studied by 
\cite{kirketal}. So, one may ask how solid this estimate is. It must be 
noted from the static model that even if the core has spent the entire 
lifetime at constant central density of $n($H$_2)=5\cdot 10^5$~cm$^{-3}$, its
chemical age is not less than 1~Myr. To make a more sounded estimate, which takes
into account the density variations, we need to rely on our dynamical 
models. It is seen from Table~\ref{tabdynpar} that, based on SP(C$^{34}$S) 
values, we can put ``error bars'' to the age estimate, so that it is 
$2\pm0.4$~Myr. But it must be kept in mind that this estimate is dependent 
on the adopted initial conditions. Therefore, it is of vital importance to 
constrain
these conditions via models of the very early stages of prestellar core 
formation and evolution. Also, the dynamical history of CB17 can be refined 
with the same analysis applied to a more complicated dynamical model, which 
we plan to do in a forthcoming study.

Alternating asymmetry pattern in the CB17 core, i.e., changing
blue-to-red asymmetry of optically thick lines and shifts of optically
thin lines, indicate that the CB17 core rotates with an estimated
azimuthal velocity of about 0.1~km s$^{-1}$. More precisely,
the model core rotates differentially with a maximum rotation velocity
of 0.13 km s$^{-1}$ at 7000~AU from the core center.
The corresponding angular momentum of the core can be estimated as
\begin{equation}
J =\frac{2}{5} M \left(R^{\rm U}\right)^{2}\Omega \approx 2\cdot 10^{56}\,{\rm g}\,{\rm cm}^{2}\,{\rm s}^{-1}.
\end{equation}
Because of our assumption that the core is observed edge-on, this is the minimum angular momentum
the core can have. Its value corresponds to the specific angular momentum of $1.6\times10^{21}$~cm$^2$
s$^{-1}$, which is too high for a single
star and somewhat higher that a corresponding value for a star with a disk
\citep[e.g.][]{ohashi1997}. On the other hand, it is close to
specific angular momenta of binary T Tau stars \citep{simon1995}.
So, it seems that CB17 is going to fragment into at least two
prestellar cores.

A possible drawback in using phenomenological descriptions is the possibility that the resulting
model may be characterized by a combination of parameters that are mutually
incompatible, like infall in a model where the thermal pressure exceeds the
gravitational pressure. In order to check if our best-fit 2D model (Table~\ref{tabdynrotpar}) is physically consistent,
we computed the gravitational force, the centrifugal force, and the pressure gradient as
functions of radius. The analysis confirms our conclusion that the core is marginally
gravitationally unstable as the gravitational force exceeds (in absolute value) 
the sum of forces, which counteract the collapse, by about 50\%. In the inner part of the core
the critical
rotation velocity that would provide enough centrifugal support against collapse
is only slightly higher that the velocity inferred from our study.
This agrees with the very low infall velocity and demonstrates that in the
inner parts of prestellar cores the rotation can be an important factor
of core dynamics. The derived kinematic structure of the CB17 core
corresponds to energy ratios $E_{\rm rot}/E_{\rm grav}\approx0.03$,
$E_{\rm therm}/E_{\rm grav}\approx0.8$, and $E_{\rm turb}/E_{\rm grav}\approx0.05$.

Our conclusions are affected to a certain degree by the  choice of the
probe molecules. All the lines we consider actually trace more the envelope of 
the core than its central region. Even optically thin lines of C$^{18}$O, H$^{13}$CO$^+$,
and C$^{34}$S are not good probes of the very center of the core because these species
are depleted there. Deuterated and N-bearing species are more promising
tracers of the inner core \citep[e.g.,][]{vandertak2005}. Inspection of
our chemical results shows that cyanopolyynes are  especially sensitive
not only to the age of an object, but also to the collapse regime. 
However, the analysis of cyanopolyyne lines with the proposed technique is not
straightforward, as it requires a more complicated approach to the radiation
transfer modeling. Another possible limitation of our model is the
neglect of  UV heating. CB17 and other similar globules are often
cometary shaped, suggestive of  some outflow from the heated surface.
Thus, UV irradiation affects not only the  chemical, but also the
thermal and the kinematic structure of the core. This is a 
promising topic for a more sophisticated dynamical model.

In this paper, we applied the `global' SP criterion, which provides an
estimate for the overall agreement between theoretical and observed spectra. 
However, as mentioned earlier, there are different features of real
spectra that can be  analyzed separately. For example, the depth of
the self-absorption dip of optically thick  lines is a good indicator of
the UV field strength, while a good measure of the core  rotation is
given by relative shifts of optically thin lines. In the latter  case
the right thing to do is to compare the line center
positions and not the overall  profile shapes. The optically thin line
intensities are sensitive to the molecule column  density, the blue-red
asymmetry depends on the infall speed, and so forth. In this sense  the
global criterion is not the best choice as it has equally bad (large)
values for a  line which is wider than the observed one (easily adjusted
with $V_{\rm turb}$; non-critical) as  well as for a line which is
shifted relative to the observed one (indicative of the  wrong rotation
model and/or the $V_{\rm LSR}$ velocity; critical). On the other hand, when
applied with  care, a SP-type criterion
provides a useful global error  control. The
global criterion is a good starting point and must be refined in a later analysis.
On the other hand, sometimes it can be complemented with  more
refined criteria that allow to study certain aspect of the model in
detail or just to  save the computational effort.

\section{Conclusions}

In this paper, we present a chemo-dynamical pilot study of the isolated 
Bok globule CB17 (L1389) based on the spectral maps of CS(2-1), C$^{34}$S(2-1),
HCO$^+$(1-0), H$^{13}$CO$^+$(1-0), and C$^{18}$O(2-1) lines. A phenomenological model of a prestellar 
core evolution combined with time-dependent chemistry and a radiative transfer 
simulation of molecular lines is used to reconstruct the 
chemical and kinematical structure of this core as well 
as to study the influence of various physical factors on molecular
line profiles. The main conclusions of the paper are:

\begin{enumerate}

\item We present a promising approach that allows to
derive the chemical and kinematic structure of a prestellar core
from its spectral maps.
We analyze both optically thick and optically thin lines, center and
off-center positions, various species and transitions etc. 
We show that even when this detailed information is available, it is not trivial to
construct a consistent core model.

\item Among the considered molecules, CS and its isotopomer C$^{34}$S turned out to be most 
sensitive species to variations in model parameters and can be used to reconstruct
the dynamical history of prestellar cores. As the effects of the age and
sticking probability ($S$) are quite similar for this molecule, it is
desirable to have more qualitative information on the $S$ value.

\item UV irradiation is an important factor affecting the chemistry and,
correspondingly, line profiles in prestellar cores even when the strength
of UV field is much smaller than the average interstellar value. The
attenuated UV field ($G=0.1$) is needed to explain
distributions of intensities and self-absorption features of the
observed CB17 spectral maps. Optically thick lines tracing the envelope
are mostly sensitive to the UV field.

\item The chemical age of the core is about 2~Myr. All the considered species (CO, HCO$^+$, and CS)
are depleted in the inner core, but the degree of depletion is still not high enough to
show up in the integrated intensity maps. This allows to classify this core as evolved
dynamically and moderately evolved chemically. The best-fit sticking probability value for the core
is $S=0.3-0.5$. This is an effective value that may be higher if we miss some important
desorption mechanism.

\item The changing asymmetry pattern of the optically thick line profiles over 
the cloud surface as well as shifts of the optically thin lines are both
indicative of a complex kinematical structure of the core.
We argue that the observed maps are reasonably well reproduced by a model
with slow infall (0.05 km s$^{-1}$), differential rotation (0.1 km s$^{-1}$),
and microturbulence (0.1 km s$^{-1}$). From the derived angular momentum,
we conclude that CB17 is likely to fragment and to form a binary (multiple) star.

\item While being artificial in nature, our phenomenological
approach allows to reveal crucial parameters that must be considered
in attempts to compare observations with results of more sophisticated
physical models based on combined MHD, chemical and radiative transfer simulations.

\end{enumerate}

\acknowledgments

We are grateful to B. Shustov, A. Tutukov (INASAN), and D. Semenov (MPIA) for fruitful discussions and to the referee for helpful comments.
This study is supported by the DFG grant ``Molecular Cloud Chemistry'' HE 1935/21-1.
YP and DW also acknowledge support from the RFRB grant 04-02-16637. DW acknowledges support from the RF President
Grant NSh-162.2003.2.


\begin{thebibliography}{}

\bibitem[Aikawa et al.(2005)]{aikawa2005} Aikawa, Y., Herbst, E., Roberts, H., Caselli, P. 2005,
\apj, 620, 330

\bibitem[Bally \& Langer(1982)]{bl1982} Bally, J., Langer, W. D. 1982, \apj, 255, 143

\bibitem[Belloche et al.(2002)]{belloche2002} Belloche, A., Andr\'e, P., Despois, D., Blinder, S. 2002, \aap, 393, 927

\bibitem[Benson et al.(1998)]{benson1998} Benson, P. J., Caselli, P., Myers, P. C. 1998, \apj, 506, 743 

\bibitem[Caselli et al.(2002)]{cas2002} Caselli, P., Benson, P. J., Myers, P. C., Tafalla, M. 2002, \apj, 572, 238

\bibitem[Chu \& Watson(1983)]{chuwatson} Chu, Y.-H., Watson, W. D. 1983, \apj, 267, 151

\bibitem[Draine(1978)]{Draine1978} Draine, B. T. 1978, \apjs, 36, 595

\bibitem[Evans(1999)]{evansaraa} Evans, N. J. II. 1999, \araa, 37, 311

\bibitem[Federman et al.(2003)]{federman} Federman, S. R., Lambert, D. L., Sheffer, Y., Cardelli, J. A.,
Andersson, B.-G., van Dishoeck, E. F., Zsarg\'o, J. 2003, \apj, 591, 986

\bibitem[Flower et al.(2005)]{flower} Flower, D. R., Pineau des For\c ets, G., Walmsley, C. M. 2005,
\aap, 436, 933

\bibitem[Geppert et al.(2004)]{geppert} Geppert, W. D., Thomas, R., Semaniak, J., Ehlerding, A., Millar, T. J. et al.
2004, \apj, 609, 459

\bibitem[Gregersen \& Evans(2000)]{greg2000} Gregersen, E. M., Evans, N. J., II. 2000, \apj, 538, 260

\bibitem[Hasegawa \& Herbst(1993)]{hh93} Hasegawa, T. I., Herbst, E. 1993, \mnras, 263, 589

\bibitem[Hogerheijde \& van der Tak(2000)]{ratran} Hogerheijde, M. R., van der Tak, F. F. S. 2000, \aap, 362, 697

\bibitem[Jijina et al.(1999)]{jijina} Jijina, J., Myers, P. C., Adams, F. C., 1999, \apjs, 125, 161 

\bibitem[Kirk et al.(2005)]{kirketal} Kirk, J. M., Ward-Thompson, D., \& Andr\'e, P. 2005, \mnras, 360, 1506

\bibitem[Kane \& Clemens(1997)]{kc1997} Kane, B. D., Clemens, D. P. 1997, \aj, 113, 1799

\bibitem[Keto \& Field(2005)]{KetoField2005}  Keto, E., Field, G. 2005, \apj, 635, 1151

\bibitem[Lada et al.(2003)]{b68pulse} Lada, Ch. J., Bergin, E. A., Alves, J. F., Huard, T. L.
2003, \apj, 586, 286


\bibitem[Lee et al.(1996)]{Lee1996} Lee, H.-H., Herbst, E., Pineau des For\c ets, G., Roueff, E.,
Le Bourlot, J. 1996, \aap, 311, 690

\bibitem[Lee et al.(2001)]{lee2001} Lee, Ch. W., Myers, P. C., Tafalla, M. 2001, \apjs, 136, 703

\bibitem[Lee et al.(2003)]{lee2003} Lee, J.-E., Evans, N. J., II, Shirley, Y. L.,
Tatematsu, K. 2003, \apj, 583, 789

\bibitem[Lee et al.(2004)]{lee2004} Lee, J.-E., Bergin, E. A., Evans, N. J., II. 2004, \apj, 617, 360

\bibitem[Lee et al.(2005)]{lee2005} Lee, J.-E., Evans, N. J., II, Bergin, E. A. 2005, \apj, 631, 351

\bibitem[Lemme et al.(1996)]{lemme1996} Lemme, C., Wilson, T. L., Tieftrunk, A. R.,
Henkel, C. 1996, \aap, 312, 585

\bibitem[Millar et al.(1997)]{umist95} Millar, T. J., Farquhar, P. R. A., Willacy, K. 1997, \aaps, 121, 139

\bibitem[Myers(2005)]{myersgen} Myers, P. 2005, \apj, 623, 280

\bibitem[Ohashi et al.(1997)]{ohashi1997} Ohashi, N., Hayashi, M., Ho, P. T. P., Momose, M.,
Tamura, M., Hirano, N., Sargent, A. I. 1997, \apj, 488, 317

\bibitem[Pavlyuchenkov \& Shustov(2004)]{pavshust} Pavlyuchenkov, Ya. N., Shustov, B. M. 2004,
Astronomy Reports, 48, 315

\bibitem[Rawlings \& Yates(2001)]{rawyates}  Rawlings, J. M. C., Yates, J. A. 2001, \mnras, 326, 1423

\bibitem[Redman et al.(2004)]{redman2004} Redman, M. P., Keto, E., Rawlings, J. M. C., Williams,
D. A. 2004, \mnras, 352, 1365

\bibitem[Sch\"oier et al.(2005)]{moldata} Sch\"oier, F. L., van der Tak, F. F. S., van Dishoeck, E. F.,
Black, J. H. 2005, \aap, 432, 369

\bibitem[Semenov et al.(2004)]{Semenov2004} Semenov, D., Wiebe, D., Henning, Th. 2004, \aap, 417, 93

\bibitem[Shematovich et al.(2003)]{shem} Shematovich, V. I., Wiebe, D. S., Shustov, B. M.,
Li, Zhi-Yun, 2003, \apj, 588, 894

\bibitem[Shirley et al.(2005)]{shirley} Shirley, Y. L., Nordhaus, M. K., Grcevich, J., M.,
Evans, N. J., II, Rawlings, J. M. C., Tatematsu, K. 2005, \apj, 632, 982

\bibitem[Simon et al.(1995)]{simon1995} Simon, M., Ghez, A. M., Leinert, Ch., Cassar, L., et al.
1995, \apj, 443, 625

\bibitem[Tafalla et al.(1998)]{taf1998} Tafalla, M., Mardones, D., Myers, P. C.,
Caselli, P., Bachiller, R., Benson, P. J. 1998, \apj, 504, 900

\bibitem[Tafalla et al.(2002)]{sysdiff} Tafalla, M., Myers, P. C., Caselli, P., Walmsley, C. M., Comito, C.
2002, \apj, 569, 815

\bibitem[Tafalla et al.(2004)]{taf2004} Tafalla, M., Myers, P. C., Caselli, P., Walmsley, C. M. 2004, \aap, 416, 191

\bibitem[Turner(1995)]{TurnerIV} Turner, B. E. 1995, \apj, 449, 635

\bibitem[Turner(1996)]{TurnerVII} Turner, B. E. 1996, \apj, 468, 694

\bibitem[Turner et al.(1997)]{TurnerVIII} Turner, B. E., Pirogov, L., Mihn, Y.C. 1997, \apj, 483, 235

\bibitem[Turner et al.(1998)]{TurnerIX} Turner, B. E., Lee, H.-H., Herbst, E. 1998, \apjs, 115, 91

\bibitem[V\'azquez-Semadeni et al.(2005)]{vslifetimes}  V\'azquez-Semadeni, E., Kim, J.,
Shadmehri, M., \& Ballesteros-Paredes, J. 2005, \apj, 618, 344

\bibitem[van der Tak et al.(2005)]{vandertak2005} van der Tak, F. F. S., Caselli, P., Ceccarelli, C. 2005, \aap, 439, 195

\bibitem[Whitworth \& Ward-Thompson(2001)]{wwt} Whitworth, A. P., Ward-Thompson, D. 2001, \apj, 547, 317

\bibitem[Wiebe et al.(2003)]{Wiebe2003} Wiebe, D., Semenov, D., Henning, Th. 2003, \aap, 399, 197

\bibitem[Williams et al.(1999)]{w99}  Williams, J. P., Myers, P. C., Wilner, D. J., di Francesco, J.
\apj, 513, L61

\end{thebibliography}
\end{document}